\def\year{2018}\relax
\documentclass[letterpaper]{article} 
\usepackage{aaai19}  
\usepackage{times}  
\usepackage{helvet}  
\usepackage{courier}  
\usepackage{url}  
\usepackage{graphicx}  
\usepackage{graphicx}
\usepackage{amsmath}
\usepackage{url}
\usepackage{textcomp}
\usepackage{booktabs}
\usepackage{todonotes}
\usepackage{multirow}
\usepackage{rotating}
\usepackage{booktabs}
\usepackage{refcount}
\usepackage{paralist}
\usepackage{array}
\usepackage{verbatim}
\usepackage{color}
\usepackage{lipsum}
\usepackage{subfig}
\usepackage{colortbl}
\usepackage{lineno,hyperref}
\usepackage{tabularx}

\RequirePackage{tcolorbox}

\begin{document}
\title{Characterizing the spread of exaggerated news content over social media}

\author{$^1$Jasabanta Patro, $^2$Sabyasachee Baruah, $^3$ Vivek Gupta,\\ \textbf{ $^4$ Monojit Choudhury, $^5$Pawan Goyal, $^6$Animesh Mukherjee}\\
$^{1,3,5,6}$IIT Kharagpur, $^2$ USC California, $^4$ MSR India \\
\{$^1$jasabantapatro, $^3$vivek\}@iitkgp.ac.in,\\ $^2$sbaruah@usc.edu,  $^4$monojitc@microsoft.com \\ \{$^5$pawang,$^6$animeshm\}@cse.iitkgp.ernet.in\\
}

\maketitle

\section{Abstract}
In this paper, we consider a dataset comprising press releases about health research from different universities in the UK along with a corresponding set of news articles. As a first step we perform an exploratory data analysis to understand how the basic information published in the scientific journals get exaggerated as they are reported in these press releases or news articles. This initial analysis shows that some news agencies exaggerate almost 60\% of the articles they publish in the health domain; more than 50\% of the press releases from certain universities are exaggerated; articles in topics like \textit{lifestyle} and \textit{childhood} are heavily exaggerated. 

Motivated by the above observation we set the central objective of this paper to investigate how exaggerated news spreads over an online social network like Twitter. We observe that the fraction of tweets sharing exaggerated news articles is higher than the ones sharing non-exaggerated tweets almost a year after the publication date of the news. In these late arriving tweets, we observe that there is a \textit{weaker presence} of words in certain LIWC categories like \textit{death words}, \textit{sexual}, \textit{sad} and \textit{negative emotion} relative to the tweets sharing non-exaggerated news; similarly, there is a \textit{stronger presence} of words in the categories like \textit{assent}, \textit{feel}, \textit{religion} and \textit{anxiety}. The LIWC analysis finally points to a remarkable observation -- these late tweets are essentially laden in words from \textit{opinion} and \textit{realize} categories which indicates that, given sufficient time, the wisdom of the crowd is actually able to tell apart the exaggerated news. As a second step we study the characteristics of the users who never or rarely post exaggerated news content and compare them with those who post exaggerated news content more frequently. We observe that the latter class of users have less retweets/mentions per tweet, have significantly more number of followers, use more slang words, less hyperbolic words and less word contractions. We also observe that the LIWC categories like `bio', `health', `body' and `negative emotion' are more pronounced in the tweets posted by the users in the latter class. As a final step we use these observations as features and automatically classify the two groups achieving an F1-score of 0.83. 
\section{Introduction}

News media has a tremendous power in shaping people's beliefs and opinions\footnote{https://bit.ly/2IAC2pD}. The Internet subculture is found to take increasing advantage of this current media ecosystem to manipulate news frames, set agendas, and propagate ideas\footnote{https://bit.ly/2tnJFJm}. Such media manipulation has significantly contributed to decreased trust of mainstream media, increased misinformation, and further radicalization\footnote{https://stanford.io/2H9kuDY}. Fake and unnecessarily exaggerated news has consequently captured worldwide interest, and the number of organized efforts dedicated solely to fact-checking has almost tripled since 2014\footnote{\protect{\url{https://bit.ly/2qbWNl4}}}.

\subsection{Exaggeration in scientific news}
Scientific news is no exception. This is especially a point of serious concern in the context of health and food related news since these are usually of wide general interest and therefore heavily consumed by the common mass~\cite{hasse2005trust}. The corresponding scientific journals and publications that appear before the news and are more complete, exhaustive and authentic account of the facts are hardly consulted by the common mass\footnote{NBC Report: \protect{\url{https://nbcnews.to/2uXjQoo}}}. The reason for this is the very abstruse and esoteric language, and academic nomenclature employed in the scientific articles, which is comprehensible often only to the health research community. Journalists and reporters, in turn, summarize the research presented in such scientific journals into a much simpler and coherent news piece, that is easily understandable by the common people.

\subsection{Forms of exaggeration}
In health related news articles, \textit{exaggeration} can take different forms, such as the \textit{transformation of correlational statements into causal statements}, \textit{change in explicitness and directness of included advice} or \textit{misrepresentation of important experimental facts}. Exaggeration can be traced to two main sources - (i) press releases made by universities~\cite{taylor2015medical,sumner2016exaggerations} and (ii) news agencies~\cite{wilson2009media,wilson2010does}. Scientists and affiliated universities, who are under tremendous pressure to make novel and interesting discoveries, make exaggerated and far-fetched claims in the press releases following a scientific publication in order to improve the chances of visibility of the research and gain increased recognition. News agencies and media houses, in addition, often exaggerate the scientific news content and the headline to make their articles more sensational and compelling. An earlier study~\cite{dumas2017poor} has concluded that journalists preferentially cover initial findings, although they are often contradicted by meta-analyses and rarely inform the public when they are later disconfirmed. Both these origins of exaggeration need to be regulated to preserve the integrity of health research. 

There have been many studies published on exaggeration~\cite{walsh2016one,sumner2014association,hasse2005trust,cullen2014mind}; however, an in-depth analysis of how these news items spread over social media is lacking. In this paper, we characterize for the first time the spread of exaggerated news items over Twitter. We also analyze the behaviour of users who never or rarely post exaggerated news content and compare them with those who post exaggerated news more frequently.


\subsection{Working definition of exaggeration}
In order to perform such a quantitative study we need to have a working definition of exaggeration. In the following we present such a definition and operationalize it later in the dataset section.

\noindent\textbf{Working definition}: In order to quantify the extent of exaggeration we measure the difference of the journal publication from the followup press release/ news articles in terms of (i) the \textit{strength of the relation between the research variables}, (ii) the \textit{explicitness and directness of the issued advice} and (iii) the \textit{identity of the sample used in experiments}. 

\subsection{Key Contributions}

Based on the above working definition, we make the following contributions in this paper.
\begin{compactitem}
\item \textit{Operationalize the definition of exaggeration}: We identify a suitably annotated dataset released by~\cite{sumner2014association} and perform an exploratory analysis. This dataset comprises detailed annotations of 462 press releases and 668 news articles, issued in 2011 by the Russel Group of Universities (20 leading UK Research Universities) in health related topics. Based on these annotations, we operationalize the definition of exaggeration. In order to ascertain the robustness of our analysis we also club close by classes to have more coarse-grained annotations. Our results do not seem to differ due to such coarse graining of the classes. 
\item \textit{Exaggeration across various dimensions}: Using the above dataset, we analyze the extent of exaggeration across various universities, media houses and health disciplines. We also demonstrate through appropriate visualizations how the strength of the causal relations and the samples used for the experiments are exaggerated in the press releases and the news articles.    
\item \textit{Propagation of the exaggerated news in Twitter}: We extensively investigate how exaggerated articles propagate in Twitter compared to non-exaggerated articles. In particular we study the tweet arrival behavior and the change in the overall linguistic construct of the tweets over time. 
\item \textit{Characterization of users spreading exaggerated news}: We compare the tweeting behaviour of users who never or rarely post exaggerated news with those posting such news content more frequently. Finally, we construct a model that can automatically classify these two groups of users using their tweet properties.
\end{compactitem}

\subsection{Result highlights}

Some of the key results that we obtain from our analysis are as follows.
\begin{compactitem}
\item From our exploratory analysis we observe the certain news agencies exaggerate up to 60\% of their articles; however, there are certain news agencies where levels of exaggeration seem pretty low. We observe that more than 50\% of press releases from certain universities are exaggerated. Finally, press releases and news articles in the \textit{lifestyle} and the \textit{childhood} disciplines are more exaggerated relative to the other sub-disciplines; on the other hand, exaggerations in the \textit{policy} sub-discipline are the least. 
\item Tweets sharing exaggerated news, on average, are more popular than those sharing non-exaggerated news.
\item The fraction of tweets sharing exaggerated news is higher compared to those sharing non-exaggerated news almost a year after the publication date of the news. 
\item If we compare the early and the late arriving tweets propagating exaggerated news, we observe stark differences in certain LIWC categories -- while the proportion of words in categories like \textit{death words}, \textit{sexual}, \textit{sad} and \textit{negative emotion} show a weaker presence, a stronger presence is observed in categories like \textit{assent}, \textit{feel}, \textit{religion} and \textit{anxiety} relative to the tweets sharing non-exaggerated news.
\item Late tweets spreading exaggerated news seem to have increased proportion of opinionated and realization tweets.
\item Users who spread exaggerated news more frequently have less retweets/mentions on average, have significantly larger number of followers, use more slang words and less hyperbolic words and contractions in their tweets compared to those users who never or rarely post exaggerated news content. Further, the LIWC categories like `bio', `health', `body' and `negative emotion' dominate in tweets posted by users who more frequently post exaggerated news content. Finally, we automatically classify these two groups of users achieving an overall F1-score of 0.83. Such classification can act as the first step in containing the spread of exaggerated content in social media early on.
\end{compactitem}

 We stress that this is the first work that investigates how exaggerated news spread in social media and shows how, given sufficient time, the wisdom of the crowd can actually tell apart the exaggerated news (increase in realization and opinion words). Further, it is also the first work that characterizes the users based on how frequently they post exaggerated content. This has the potential to help in containing the spread of such content in social media.


\section{Related work}

Investigating sensationalism in scientific news has been an area of research interest for quite some time although rigorous quantitative  measurement studies are missing. In the following, we present a brief summary of some of the existing studies.

In~\shortcite{hoffman2016automatically}, the authors defined two measures -- scientific quality and sensationalism, and devised an entropy model to learn the same for news paper articles. The annotations were crowd sourced through a questionnaire. Schwartz et al.~\shortcite{schwartz2012influence} investigated whether the quality of the news articles can be influenced by the quality of press releases. Yavchitz et al.~\shortcite{yavchitz2012misrepresentation} coined a term called ``spin'' as specific reporting strategies (intentional or unintentional) emphasizing the beneficial effect of the experimental treatment, and investigated the same for 498 press release articles. Woloshin et al.~\shortcite{woloshin2009press} investigated the works of 10 medical centers at each extreme of U.S. News \& World Report's rankings for medical research, and concluded that press releases from academic medical centers often promote research that has uncertain relevance to human health and do not provide key facts or acknowledge important limitations. Selvaraj et al.~\shortcite{selvaraj2014media} studied which medical research was most likely to be covered in media, and concluded that observational studies had the highest frequency. Brechman et al.~\shortcite{brechman2009lost} examined the presentation of genetic research relating to cancer outcomes and behaviors in both the press release and its subsequent news coverage. 
Adams et al.~\shortcite{adams2017readers} studied how people understand scientific expressions used in news headlines. In~\shortcite{sumner2014association} the authors study the association between exaggeration in health related science news and academic press releases. In this study, the authors report the overall levels of exaggeration in the claims made in the press releases, in the advice statements made and the samples used for the experiments. In this paper, we use the same dataset made public by the authors as well as the annotations that they use to mark exaggerated content. However, instead of analyzing the overall levels of exaggeration, we launch a more \textit{micro-scale} study and observe levels of exaggeration \textit{university-wise}, \textit{news agency-wise} as well as \textit{sub-discipline-wise}. We further investigate for the first time how the exaggerated content spreads in an online social network like Twitter. 
\section{Dataset}\label{data}
We use the publicly available dataset\footnote{Dataset: \url{https://bit.ly/2qc86tk}} released by~\cite{sumner2014association} for our experiments. This dataset contains detailed annotations of 462 journals, corresponding 462 press releases and 668 news articles, issued in 2011 by the Russel Group of Universities (20 leading UK Research Universities) in health related topics, for three distinct categories - \textit{strength of statement}, \textit{advice} and \textit{sample}. Every individual press release is a followup of a journal paper; we assume this journal paper to be the reference for our analysis (as has also been assumed in~\cite{sumner2014association}). Every press release in turn, can be discussed by some news reports. In our dataset, 230 out of 462 journals and press releases have at least one news article coverage. The data set primarily spans over eight disciplines. The number of press releases and articles in each of these disciplines are shown in Table~\ref{Table:DisciplineDist}. 

\begin{table}
\centering
\small
\begin{tabular}{|c|c|c|}
\hline
Discipline & press releases & news articles \\ \hline
Life-style & 70 & 121  \\
Mental Health & 14 & 14 \\ 
Childhood & 43 & 58  \\
Treatment & 61 & 108 \\ 
Observational Identification & 203 & 282  \\
Policy & 29 & 30 \\ 
Ageing & 3 & 4  \\
Physical Disease & 38 & 49 \\
Not Mentioned & 1 & 2 \\
\hline 
\end{tabular}%
\vspace{0.2cm}
\caption{Number of press releases and articles in each discipline.}
\label{Table:DisciplineDist}
\vspace{-0.8cm}
\end{table}


In order to operationalize the definition of exaggeration one needs to be aware of the statements appearing in the different documents (journal papers, press releases or news articles). This is done as follows.

\subsection{Strength of statement}  Health related research experiments involve various compounds and quantities that are measured and controlled. \textit{Independent variables} $(IV)$ refer to those variables that are varied in the experiment while \textit{dependent variables} $(DV)$ refer to those variables that get affected and are measured. This could be thought of as analogous to the subject and the object in a natural language sentence. For example, consider the sentence \textit{Wine causes cancer}. Here `wine' is the $IV$ and `cancer' is the $DV$. We may have multiple $IV$s and $DV$s for a research study, and they can be determined by carefully reading through the journal article. The strength of statement is defined as a numerical measure of the strength of the causal relationship of the statement that relates an $IV$ and a $DV$. In~\cite{sumner2014association}, this strength has been quantized into seven different levels that are noted below. 

\begin{enumerate}
\itemsep1pt\relax
\item	No relationship mentioned - No relationship is mentioned, e.g., ``wine and cancer''.
\item Statement of no relationship - Explicitly stating that there is no relationship, e.g., ``wine \textbf{does not cause} cancer''.
\item Statement of correlation - $IV$ and $DV$ are associated, but causation cannot be explicitly asserted, e.g.``wine \textbf{is associated with} cancer''.
\item Ambiguous statement of relationship - It is unclear what the strength of relationship of this statement is, e.g.,``wine \textbf{is linked to} cancer". This could mean that ``wine \textbf{causes} cancer'', or that ``wine \textbf{is associated with} cancer'' - either would be applicable.
\item Conditional statement of causation - Causal statements show that the $IV$ directly changes the $DV$. Conditional causal statements carry an element of doubt in them, e.g., ``wine \textbf{might cause} cancer''.
\item Statement of ``can" - The word ``can" is unique as a statement of relationship in that it implies that the $IV$ has the potential to directly change the $DV$, e.g., ``wine \textbf{can cause} cancer''. Therefore it is stronger than any conditional statement of causation.
\item Statement of causation - The strongest are the statements of causation, e.g., ``wine \textbf{causes} cancer''. This statement says that the $IV$ definitely and directly alters the $DV$.
\end{enumerate}

The strength of statement level (1--7) of an entire document (e.g., a journal paper, a press release or a news article) is defined by the level of the strongest statement relating an $IV$ and a $DV$ appearing in that document. We shall refer to this strongest statement as the \textit{first statement} throughout the rest of the paper\footnote{Note that there can be other statements apart from the first statement in the article which are not considered in this study.}.

While~\cite{sumner2014association} provides the dataset already marked by one of the above seven quantization level for each journal, press release and news article, thus allowing for analysis of the exaggerated content, it is difficult to ascertain the robustness of the results obtained. This is primarily because some of the quantization levels seem to be too close. In order to test the robustness of the results that we present in the subsequent sections, we also club the above quantizations into more coarse-grained labels. Essentially, we consider a 4-class and a 2-class quantization. For the 4-class we map the above seven quantizations as follows: $(1, 2) \rightarrow 1$, $(3, 4) \rightarrow 2$, $(5, 6) \rightarrow 3$ and $7 \rightarrow 4$. For the two class we map as follows: $(1, 2, 3, 4, 5, 6) \rightarrow 1$ and $7 \rightarrow 2$.

\subsection{Advice}
Health related research often conclude with a statement of advice, given explicitly or implicitly, to the reader usually recommending what one should or should not do/ eat. Often such advice comes in the form of a quote from the lead researcher of the scientific article. \textit{Advice} is quantized at four different levels based on the explicitness and directness in the lines of~\cite{sumner2014association}. We enumerate these levels below in their ascending order of strength.

\begin{enumerate}
\itemsep1pt\relax
\item No advice - If no advice is given.
\item Implicit advice - If advice is implicit and hints that certain behavior needs to be changed, e.g., ``Drinking wine is harmful because it causes cancer''. This implies that public should not drink wine but does not directly or explicitly state it.
\item Explicit and indirect advice - Advice is explicitly stated but is not directly addressed to the public, e.g., ``Drinking wine causes cancer, therefore governments should advice their citizens not to drink wine''.
\item Explicit and direct advice - Advice is explicitly stated and is directly addressed to the public, e.g., ``We should not drink wine because it causes cancer''.
\end{enumerate} 

Once again, the advice level (1--4) of a document corresponds to the level of the strongest advice statement appearing in the document. Note that these annotations are again already marked in the dataset made available by~\cite{sumner2014association} for each journal, press release and news article.

\subsection{Sample}
Research experiments can be conducted on humans, mice, bees, primates, rodents or even on diseased cells. In~\cite{sumner2014association} the authors make nine different classes of \textit{sample}. However, an initial analysis indicated that the data becomes too sparse if one considers nine classes. Thus for our purpose, we simplify this classification by putting all human subject related research studies in the \textbf{human} class, and the rest non human subject based experiments into the \textbf{non-human} class. We assign the sample class \textbf{human} to a document if the research explicitly or implicitly states that human subjects were used or if the study was intended for humans. Otherwise, we assign the class \textbf{non-human} to the document.

In the next section, we operationalize the definition of \textit{exaggeration} by comparing the level of the strength of statement, the level of the advice and the class of the sample.


\section{Operational definition of exaggeration}\label{defn}
We define \textcolor{black}{three different variants for quantification of} exaggeration by considering the journal paper as always the reference non-exaggerated document. \textcolor{black}{ These three variants differ based on how we quantize the strength of statement (7, 4 or 2).} Suppose $d$ is a document (a press release or a news article), with strength of statement level $s_d$, advice level $a_d$ and sample class $x_d$. Also suppose $j$ is the reference journal with strength of statement level $s_j$, advice level $a_j$ and sample class $x_j$, \textcolor{black}{where $a_d, a_j \in \{1,2,3,4\}$ and $x_d, x_j \in \{\textbf{human},\textbf{non-human}\}$. The values of $s_d, s_j$ are defined as per the following variants of quantization:}
\textcolor{black}{ \\
\noindent \textbf{7-Class:} Here we take strength levels as have been defined in~\cite{sumner2014association}. So $s_d, s_j \in \{1,2,3,4,5,6,7\}$. \\
\noindent \textbf{4-Class:} As discussed earlier, here we shrink the seven classes to four classes. The mapping from the original seven class to the four class is as follows.  $(1, 2) \rightarrow 1$, $(3, 4) \rightarrow 2$, $(5, 6) \rightarrow 3$ and $7 \rightarrow 4$. So $s_d, s_j \in \{1,2,3,4\}$.\\
\noindent \textbf{2-Class:} For the two class we map the original seven classes as follows: $(1, 2, 3, 4, 5, 6) \rightarrow 1$ and $7 \rightarrow 2$. So $s_d, s_j \in \{1,2\}$.}    

\noindent Exaggeration therefore is defined as 
\begin{enumerate}
\itemsep1pt\relax
\item $d$ is exaggerated in strength of statement if $s_d > s_j$.
\item $d$ is exaggerated in advice if $a_d > a_j$.
\item $d$ is exaggerated in sample if $x_d = \textbf{human}$ and $x_j = \textbf{non-human}$.
\end{enumerate}

Overall, the document $d$ is exaggerated with respect to $j$ if $d$ is exaggerated in strength of statement \textbf{or} in advice \textbf{or} in sample. 

\section{Exploratory analysis}\label{sec:eval}

In this section we present an extensive measurement study of different forms of exaggeration in the strength of statement, advice and sample. In specific, we investigate the extent of exaggeration across various (i) news agencies, (ii) universities as well as (iii) disciplines. 

\subsection{Exaggeration by news agencies}

\begin{figure*}[!t]
\centering    
\subfloat[7-Class]{\label{Figure:News7Class}\includegraphics[width=0.5\textwidth]{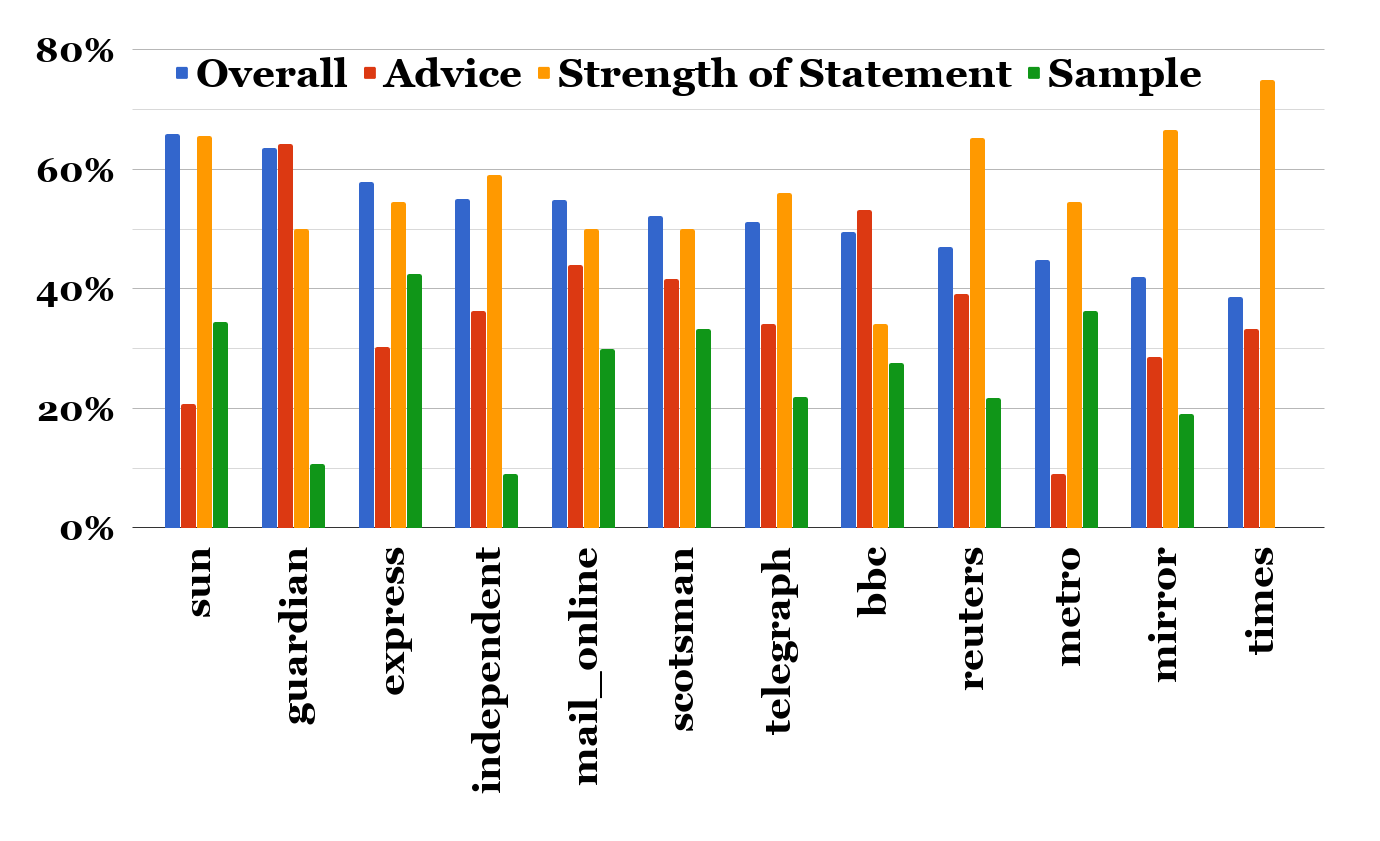}}
\subfloat[2-Class]{\label{Figure:News2Class}\includegraphics[width=0.5\textwidth]{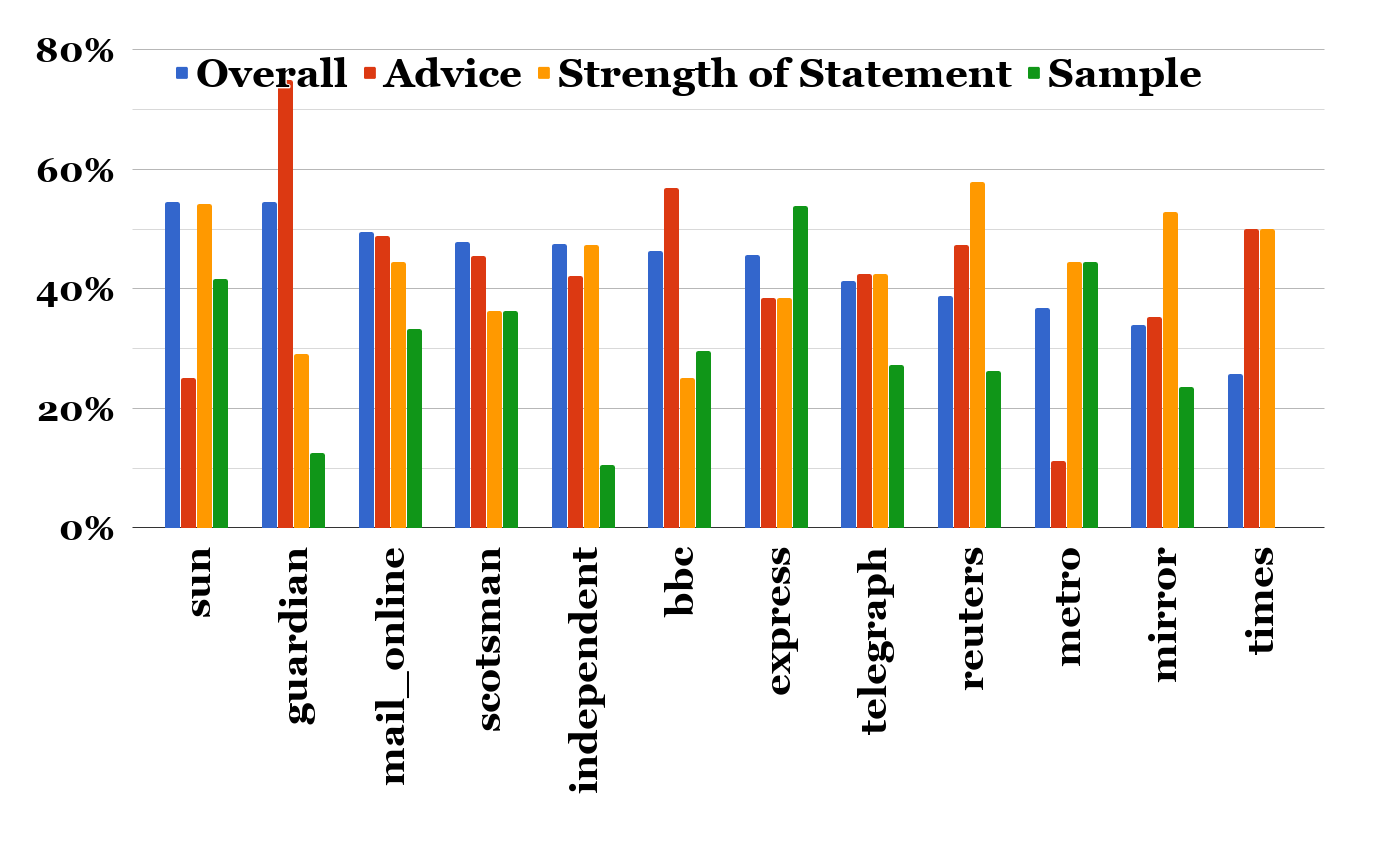}} 
\caption{Exaggeration percentages for news media in descending order of the overall exaggeration percentage.}
\label{News_Source}
\vspace{-0.3cm}
\end{figure*}

Our dataset comprises news articles from 21 different media outlets including some of the leaders like \textit{BBC}, \textit{Reuters}, \textit{Guardian}, \textit{Telegraph} and \textit{Daily Mail}. In Figure~\ref{News_Source}, we report the overall as well as the category level exaggeration values for some of the top media outlets. For strength of statement, we show the 7-class and 2-class variants only (the results of the 4-class variant are very similar and therefore not shown). We note some of the key observations below.

\begin{compactitem} 
\item Certain media outlets seem to overall exaggerate in more than 60\% (50\%) of the articles they publish considering 7-class (2-class) variants in the strength of statement. Nevertheless, there are some outlets that keep the exaggeration level very low.
\item An interesting observation is that the news agencies exhibiting higher overall levels of exaggeration as per the 7-class as well as the 2-class variants are very similar. For the 4-class too, this trend prevails. 
\item Most media outlets exaggerate heavily in their strength of statements. However, for certain media houses one also finds exaggerations in advice and sample levels.
\end{compactitem}

\subsection{Exaggeration by universities}
\begin{figure*}
\centering    
\subfloat[7-Class]{\label{Figure:Univ7Class}\includegraphics[width=0.5\textwidth]{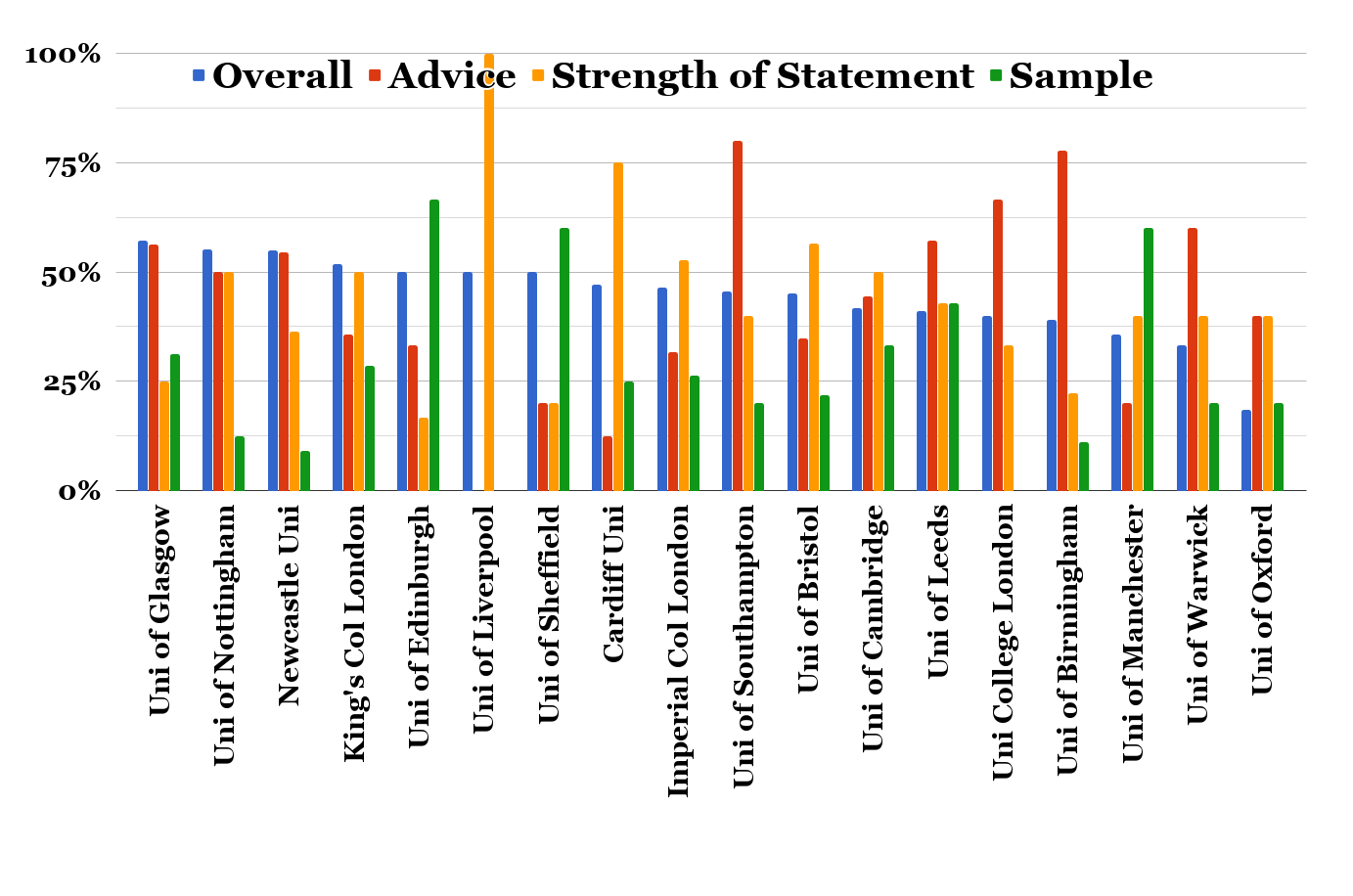}} 
\subfloat[2-Class]{\label{Figure:Univ2Class}\includegraphics[width=0.5\textwidth]{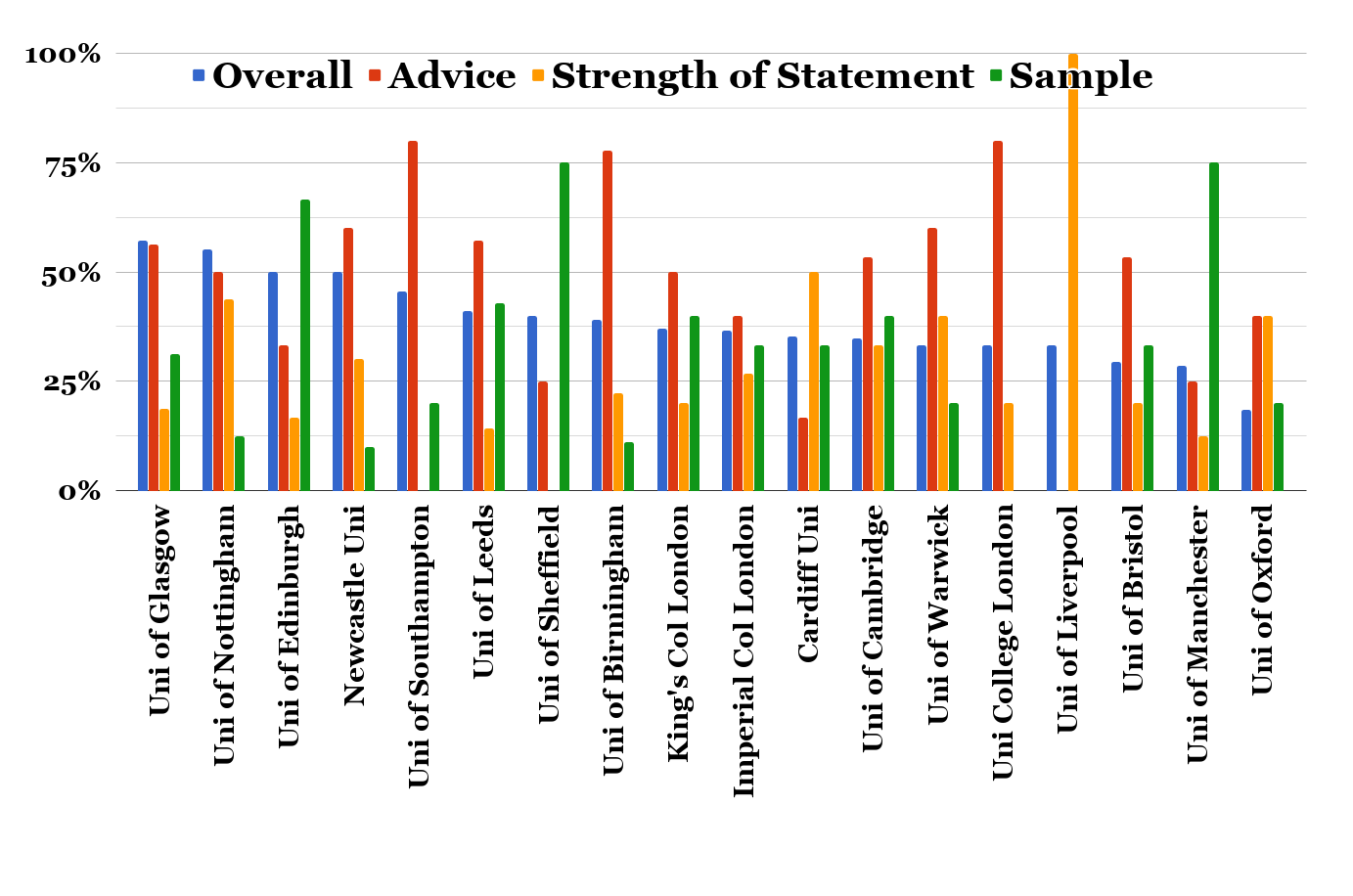}}
\caption{Exaggeration percentages among press releases of universities in descending order of overall exaggeration percentage.}
\label{University}
\vspace{-0.4cm}
\end{figure*}

We have 20 research universities in our dataset. In Figure~\ref{University}, we report the overall and the category level exaggeration percentages among the press releases made by these universities \textcolor{black}{considering the 7-class and the 2-class variants of strength of statement.} The trends for the 4-class variant are again very similar and therefore not shown. We enumerate some of the important observations below.
\begin{itemize}
\item Like media houses, certain universities seem to exaggerate their press releases heavily. However, there are certain universities that rarely exaggerate their press releases. 
\item Some universities exaggerate mostly in the sample category. This means that results on non-human subjects are reported as being done on human subjects.
\item Many universities tend to exaggerate in the advice category. There are also some universities that exaggerate in the strength of statements in \textit{all} the six press releases present in our dataset.
\item For both the 7-class and the 2-class classification of the strength of statements, similar universities come at the top in overall levels of exaggeration.
\end{itemize}

\if{0}
\noindent\textbf{Insights}: Some of the insights obtained from this analysis are - (i) while many universities in our dataset tend to exaggerate their press releases, there are some that have been able to keep away from this (mal)practice\footnote{\url{https://go.nature.com/2JsMUr3}}; (ii) a lot of health related advice seem to be, in general, accompanied with university press releases possibly to attract more media attention. We believe that such exaggeration practices should be taken into account by the government (and other) agencies responsible for ranking universities\footnote{\url{https://go.nature.com/2m73vZI}}.\fi

\subsection{Exaggeration across disciplines}

\begin{figure*} [h]
\centering    
\subfloat[7-Class]{\label{Figure:Disc7Class}\includegraphics[width=0.5\textwidth]{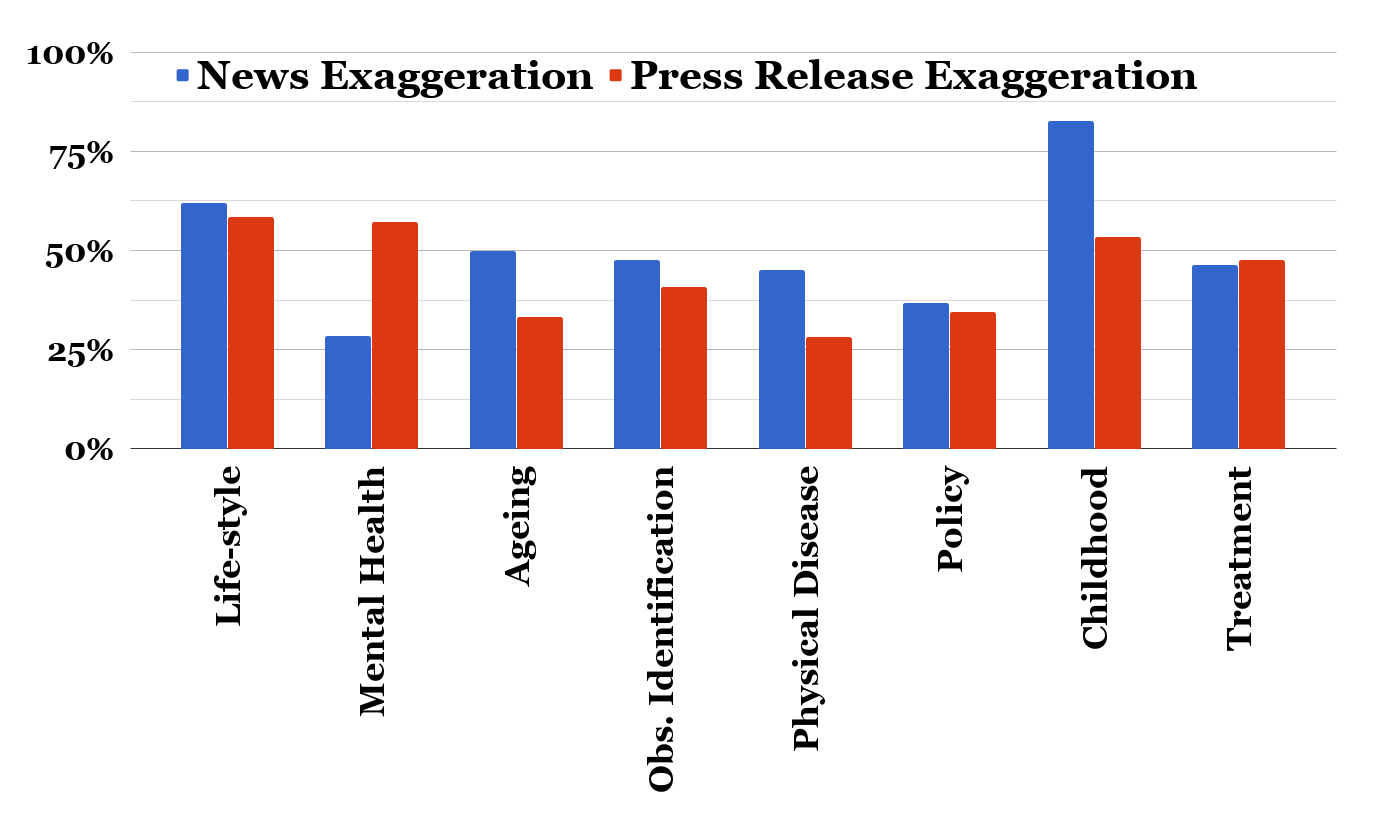}} 
\subfloat[2-Class]{\label{Figure:Disc2Class}\includegraphics[width=0.5\textwidth]{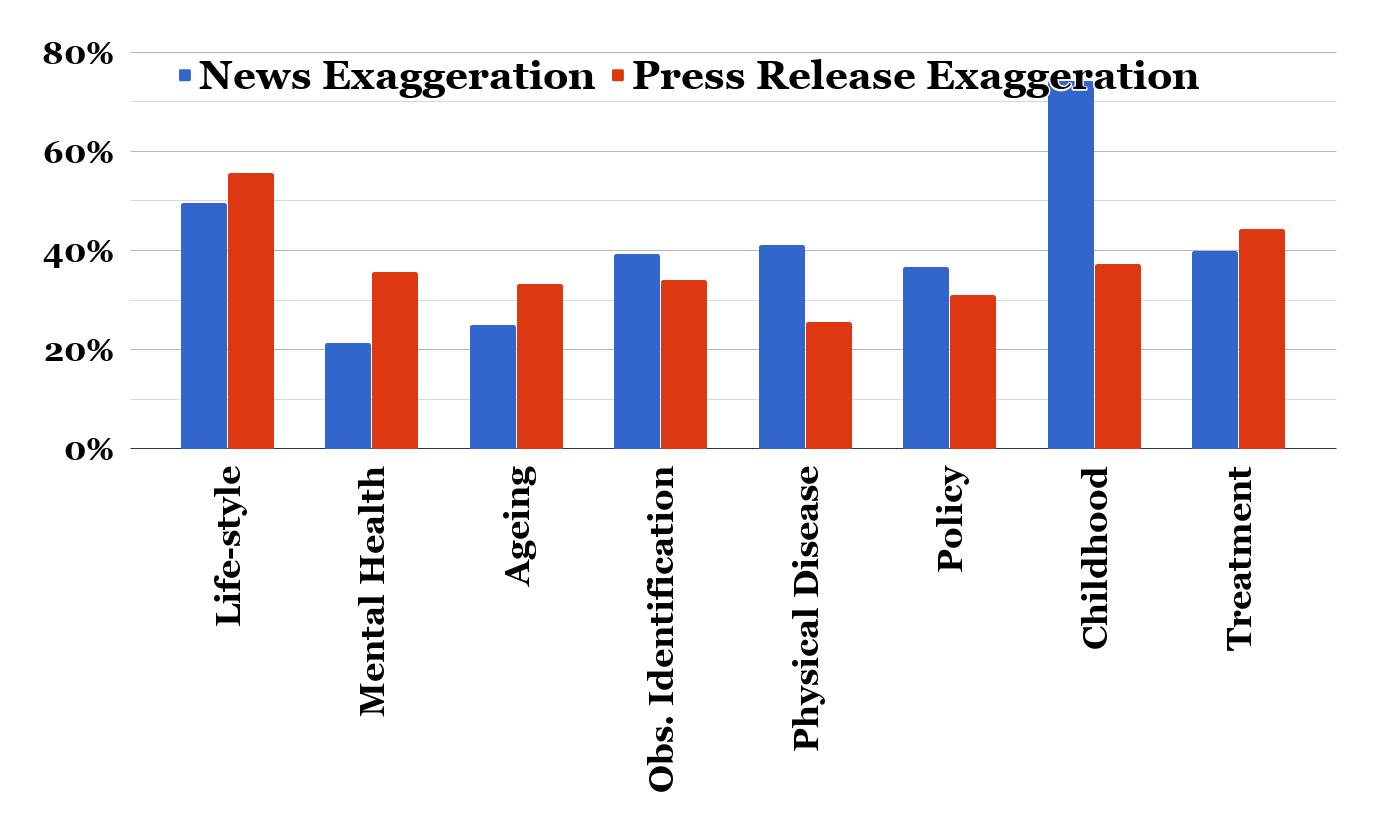}}
\caption{Exaggeration percentages in news articles and press releases in different disciplines.}
\label{Discipline}
\vspace{-0.5cm}
\end{figure*}

Each journal article can be categorized into a discipline. There are, in all, eight disciplines - \textit{childhood}, \textit{lifestyle}, \textit{treatment}, \textit{policy}, \textit{observational identification}, \textit{ageing}, \textit{mental health} and \textit{physical disease}. This categorization also extends to the followup press release and the news articles. We examine the extent of exaggeration in each of these disciplines in Figure~\ref{Discipline}. Some of the important observations here are provided below.
\begin{compactitem}
\item The maximum overall exaggeration appears in the \textit{childhood} and the \textit{lifestyle} disciplines. This is true for the 7-class, the 4-class as well as the 2-class variants of strength of statements used in the definition of overall exaggeration. 
\item In general, exaggeration percentages are more in news articles than in the press releases. However, in some disciplines like \textit{mental health} the exaggeration percentage in press releases is double or less compared to the percentage in news articles.
\end{compactitem}


\if{0}
\subsection{Variations due to exaggeration}
In this section, we study the interesting microscopic variations in the level of the strength of statement and the class of the sample.

\subsubsection{Variation in strength of statement level}
Recall that the strength of relation between $IV$ and $DV$ in a statement is defined in purely syntactic terms. Therefore, the word \textbf{can} and other synonymous clauses render the relation a strength of six whereas the word \textbf{might} renders a strength of five (7-class). In Figures~\ref{Alluvial_Statement_PR4} and~\ref{Alluvial_Statement_PR2}, we show how the strength of a relation between an $IV$-$DV$ pair gets changed from the journal to the corresponding press release for the 4-class and the 2-class variants, respectively. Similarly, in Figures~\ref{Alluvial_Statement_NA4} and ~~\ref{Alluvial_Statement_NA2}, we illustrate how the strength of the relation between an $IV$-$DV$ pair gets changed from the journal to the followup news articles for the 4-class and the 2-class variants, respectively. Some of the interesting observations are noted below.
\begin{itemize}
\item A major fraction of the correlational statements reported in the journals get (mis)represented as causal statements in the corresponding press releases.
\item In many press releases, causal connections are drawn between $IV$-$DV$ pairs that are no way related in the source journals.
\item In case of news articles, correlational and ambiguous statements in the journal get (mis)represented as either conditional causal statements (might) or unconditional causal statements. 
\end{itemize}
\begin{figure} [h]
\subfloat[Journal to Press Release (4-class)]{\label{Alluvial_Statement_PR4}\includegraphics[width=0.25\textwidth, height=4cm]{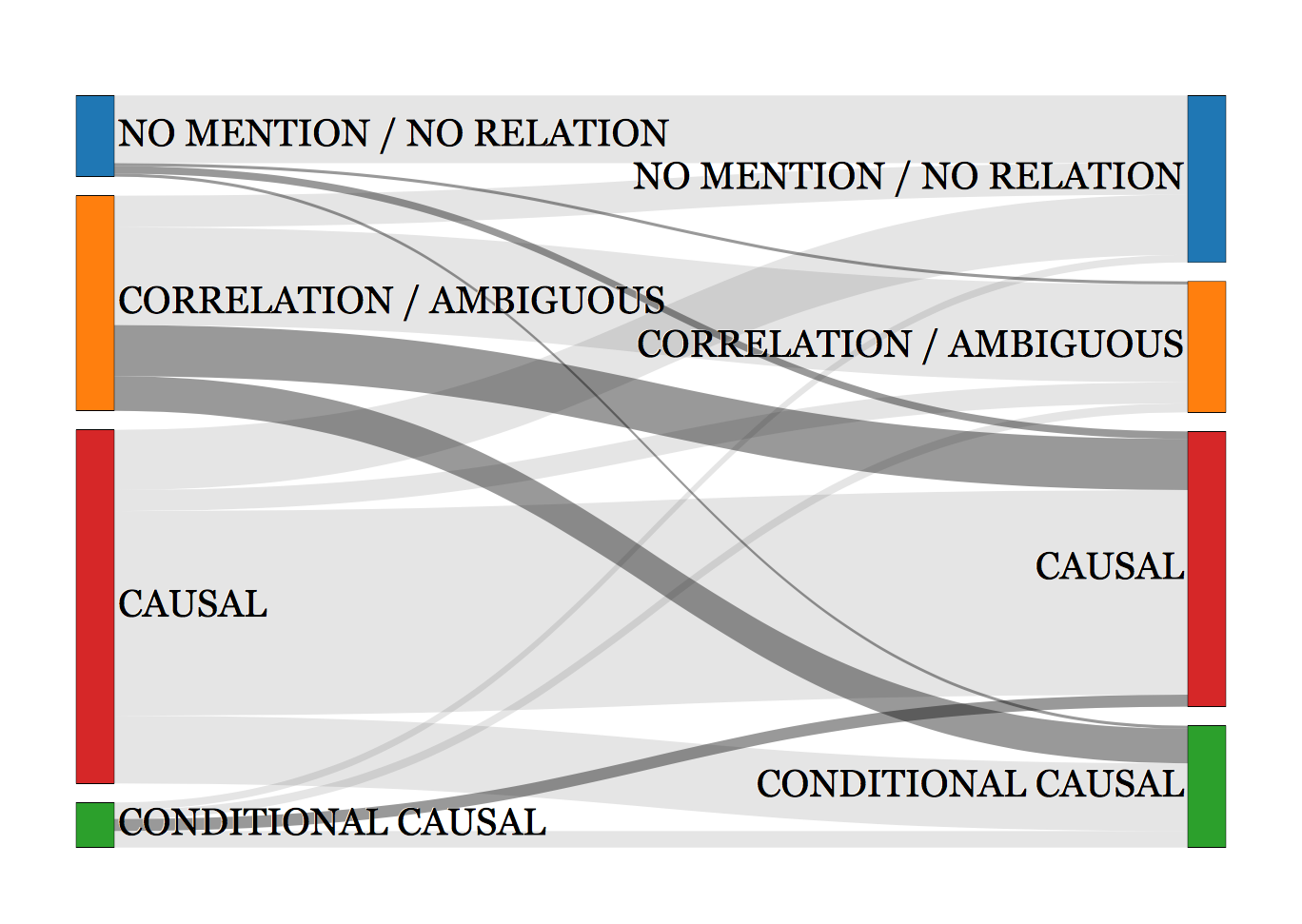}}
\subfloat[Journal to Press Release (2-class)]{\label{Alluvial_Statement_PR2}\includegraphics[width=0.25\textwidth, height=4cm]{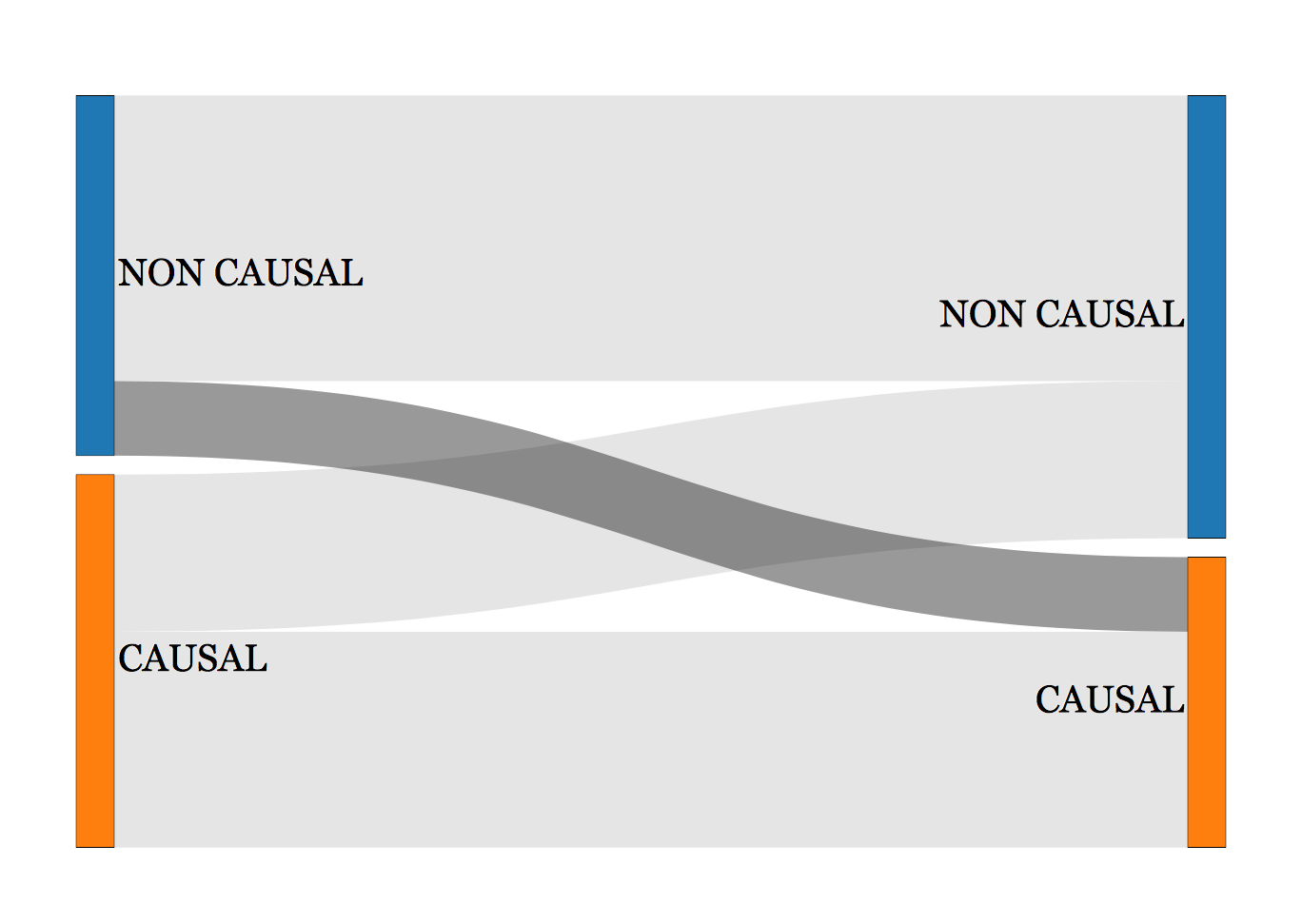}} \\

\subfloat[Journal to News Article (4-class)]{\label{Alluvial_Statement_NA4}\includegraphics[width=0.25\textwidth, height=4cm]{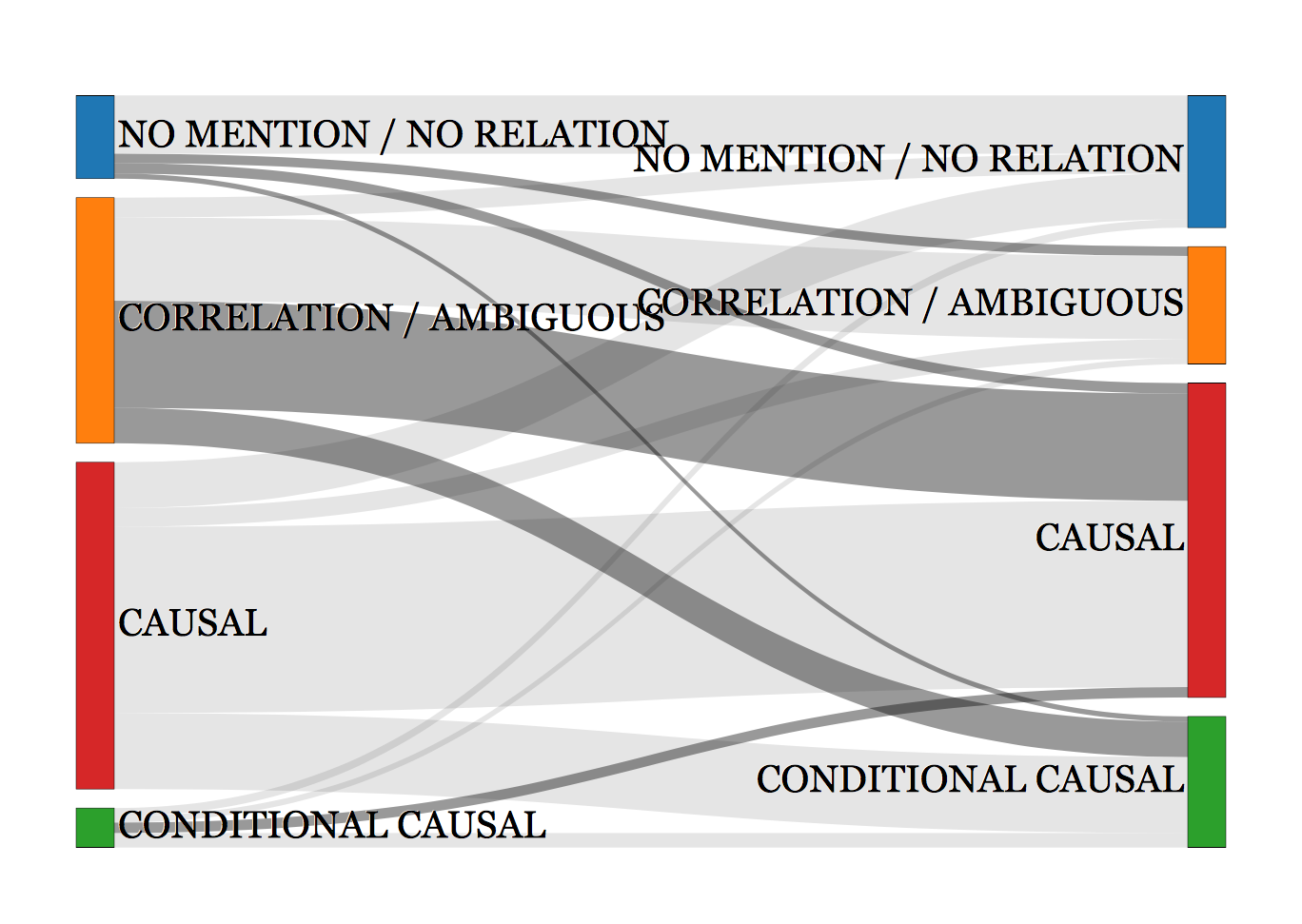}}
\subfloat[Journal to News Article (2-class)]{\label{Alluvial_Statement_NA2}\includegraphics[width=0.25\textwidth, height=4cm]{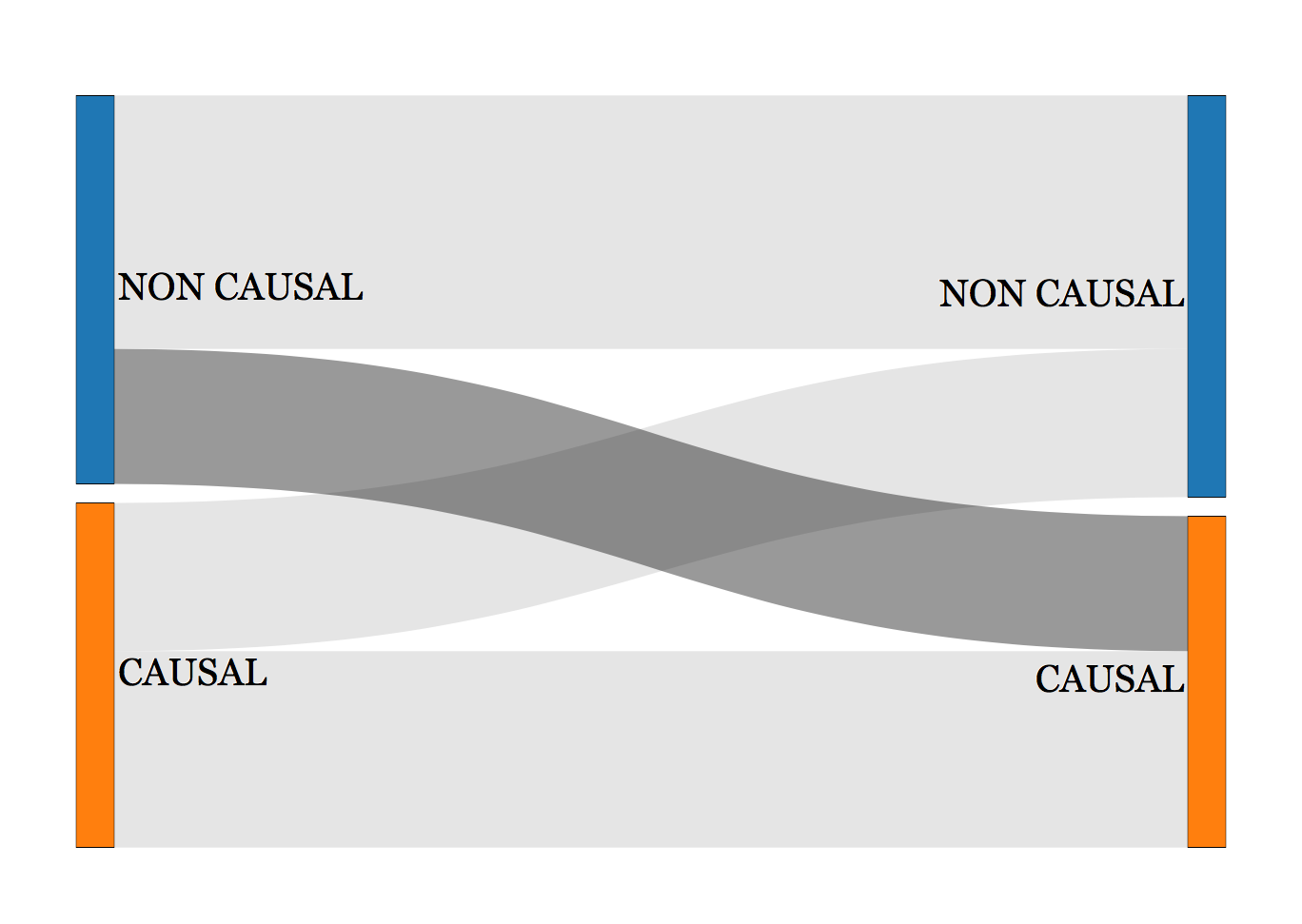}}
\caption{Change in the level of the strength of statement. The most important trends are darkened.}
\vspace{-0.4cm}
\end{figure}



\subsubsection{Variation in sample class}
Recall that we have two classes of sample -- \textbf{human} and \textbf{non-human}. The \textbf{non-human} class can further be divided into the following -- rodents, primates, cells, computer simulations etc. In Figures~\ref{Alluvial_Sample_PR} and~\ref{Alluvial_Sample_NA} we respectively illustrate how the sample class is changed in the press releases and the news articles from the source journal articles. In both cases we observe that research conducted originally on rodents (mice, rats etc.) and cells (cancer cells, \textit{in vivo} etc.) are misreported as being performed on human subjects in the press releases/news articles.
\begin{figure}[h]
\centering    
\subfloat[Journal to Press Release.]{\label{Alluvial_Sample_PR}\includegraphics[width=0.25\textwidth, height=4cm]{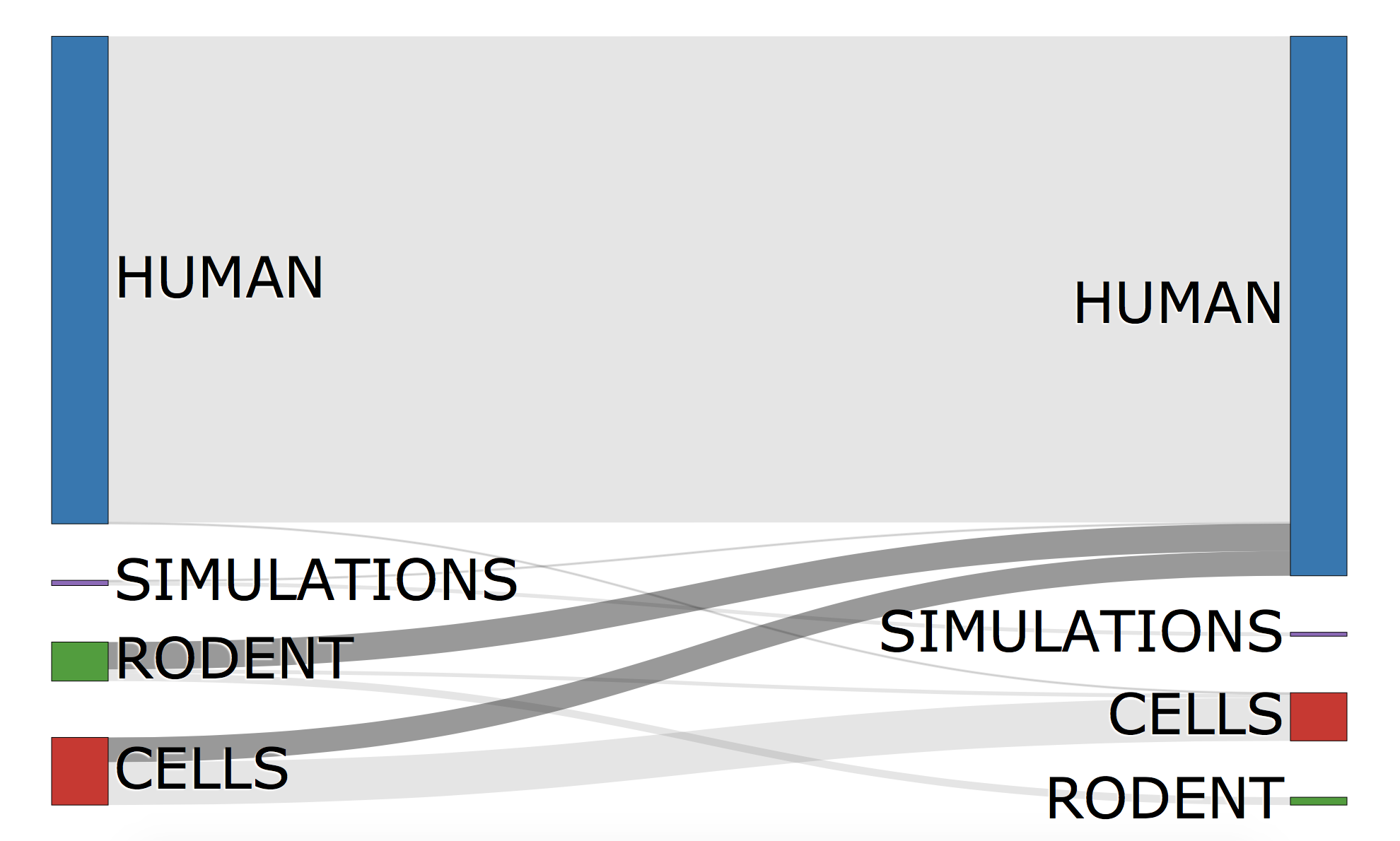}}
\subfloat[Journal to News Article.]{\label{Alluvial_Sample_NA}\includegraphics[width=0.25\textwidth, height=4cm]{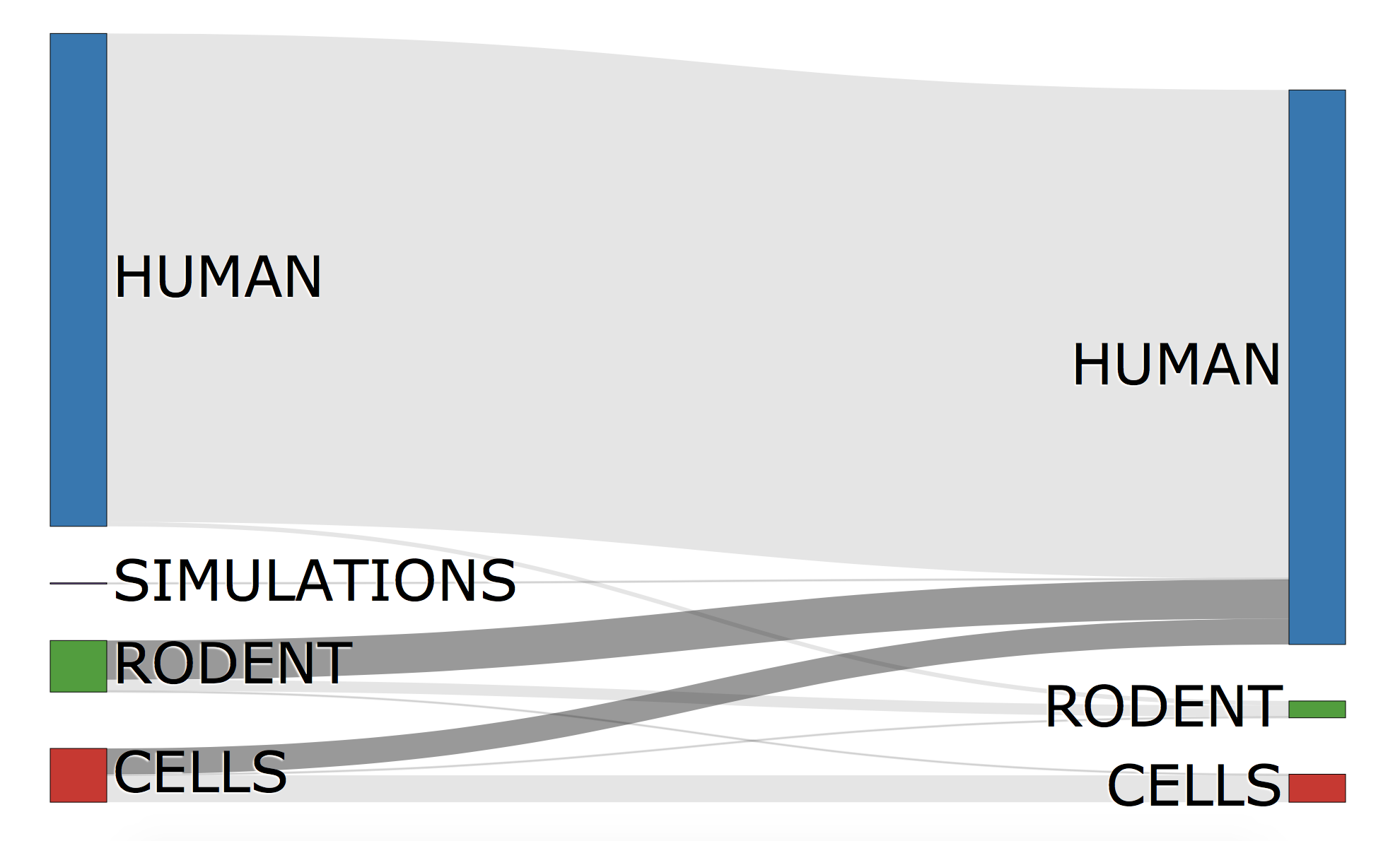}}
\caption{Change in the sample class. The most important trends are darkened.}
\vspace{-0.6cm}
\end{figure}

\subsection{Analysis of the text connecting IV and DV}

We further investigate the text of the statements that connect the $IV$ and $DV$ of the first statements. Figure~\ref{phrasecloud} shows the phrases in the first statements when we consider the 4-class variant. As expected, the strength levels 1 and 2 show correlative words/phrases like ``link to'', ``associate'', ``be with'' etc. while levels 3 and 4 mostly contain words like ``can'', ``could'' etc. or fully causal words like ``reduce'', ``improve'', ``increase'' etc.

\begin{figure*} [!htbp]
\centering    
\subfloat[Strength level 1]{\label{Wordcloud_0}\includegraphics[width=0.25\textwidth, height=3cm]{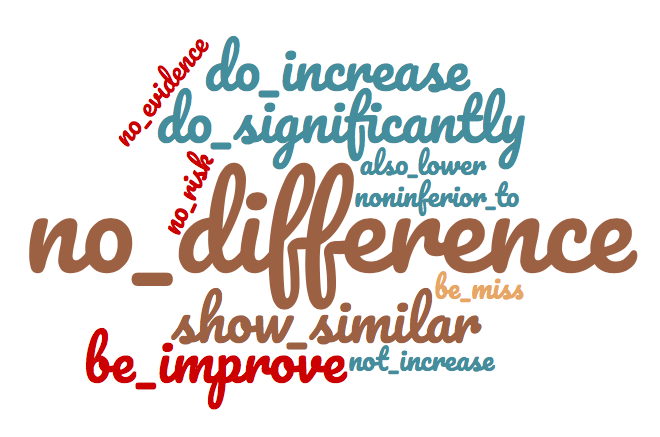}}
\subfloat[Strength level 2]
{\label{Wordcloud_1}\includegraphics[width=0.25\textwidth, height=3cm]{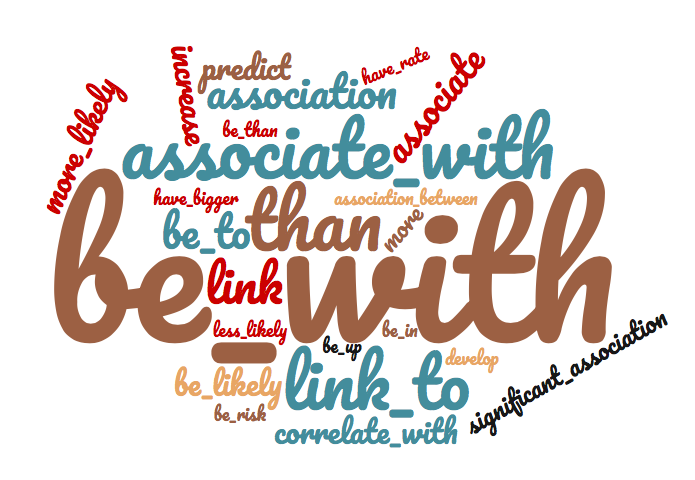}}
\subfloat[Strength level 3]
{\label{Wordcloud_2}\includegraphics[width=0.25\textwidth, height=3cm]{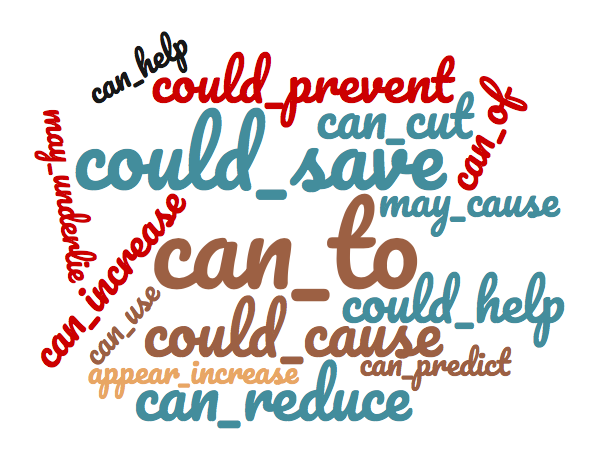}}
\subfloat[Strength level 4]
{\label{Wordcloud_3}\includegraphics[width=0.25\textwidth, height=3cm]{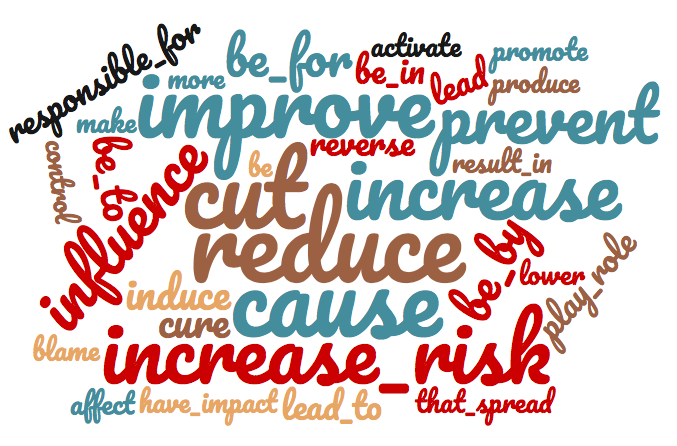}}
\caption{\label{phrasecloud} Phrase clouds of the statements connecting $IV$ and $DV$ of the first statement.}
\vspace{-0.6cm}
\end{figure*}
\fi

\section{Spread of news articles in Twitter}
In this section, we investigate how social media discussion of the news articles in our dataset propagates on Twitter. We believe that this a unique, important and novel contribution of our paper. In this context, we shall denote tweets discussing exaggerated news articles by $tweet_{EN}$, and those discussing non-exaggerated news articles by $tweet_{NEN}$.

\noindent\textbf{Collection process}: We collect the headline of each news article, and use it to search for the corresponding webpage. From the webpage we record the headline, url, text and links displayed when we click on the `tweet share' button (if it exists), date, time and the name of the author. We might not find any online version of our news article sometimes, especially if the news article originally appeared in the print media. In this case, we exclude the news article from further investigation. With the aforementioned metadata for each news article, we search and collect tweets. To this purpose, we use the python library \href{https://github.com/Jefferson-Henrique/GetOldTweets-python}{GetOldTweets} from GitHub\footnote{The official Twitter python API \textit{tweepy} could not meet our requirement because our news articles date from 2011.}. From the tweets collected, we retain only those that were posted after the news article was published. To avoid spam, we further remove tweets that do not contain any url that link to the news article's webpage. The total number of tweets that we finally collect is 14,686; the number of tweets in the $tweet_{EN}$ and $tweet_{NEN}$ \textcolor{black}{considering the three class variants of strength of statement are shown in Table~\ref{Table:TweetDist}. As the table shows, the tweet distribution is roughly balanced across the $tweet_{EN}$ and $tweet_{NEN}$ categories for all the three variants.}

\begin{table}[!htbp]
\centering
\begin{tabular}{|c|c|c|c|c|}
\hline
classification & $|tweet_{EN}|$ & $|tweet_{NEN}|$ \\ \hline
7 class & 7,689 & 6,997 \\
4 class & 7,255 & 7,431  \\
2 class & 6,952 & 7,734  \\ \hline
\end{tabular}%
\vspace{0.2cm}
\caption{Number of tweets in our dataset as per the three variants.}
\label{Table:TweetDist}
\vspace{-0.3cm}
\end{table}

In Table~\ref{Table:Tweet} we note some of the basic properties of these \textcolor{black}{two types of tweets for all the three class variants}. Fraction of likes, retweets and mentions for the $tweet_{EN}$ category seem to indicate that the tweets that propagate exaggerated news are usually more popular.

\subsection{Arrival time}
\begin{figure}[!h]
\centering    
\subfloat[7 - Class]{\label{Figure:7ClassArrival} \includegraphics[width=0.25\textwidth]{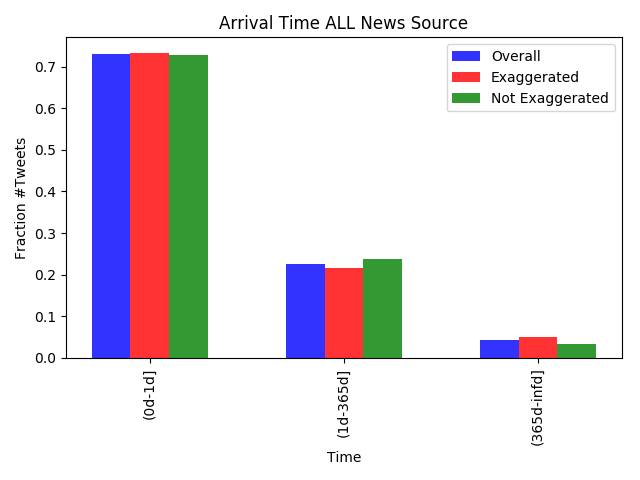}} 
\subfloat[4 - Class]{\label{Figure:4ClassArrival} \includegraphics[width=0.25\textwidth]{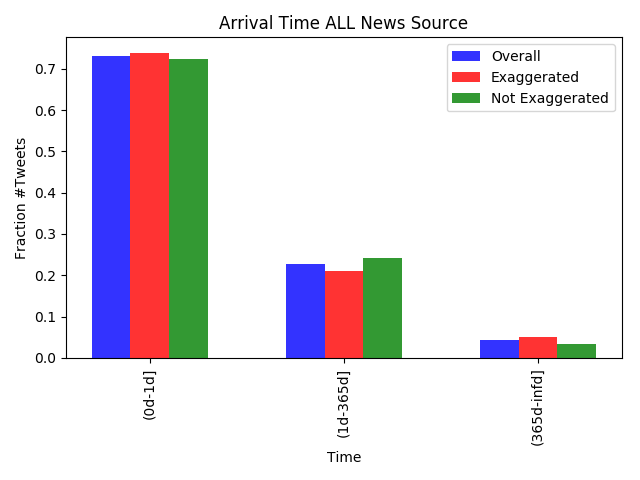}} 
\caption{Arrival time distribution of tweets.}
\label{Figure:Arrival}
\end{figure}

In order to understand the dynamics of the propagation we study the arrival time distributions of the tweets as follows. In particular, in Figure~\ref{Figure:Arrival}, we plot the fraction of tweets in each of the exaggerated and non-exaggerated groups as well as the overall fraction that arrive within one day, between one day and one year and after one year from the date of publication of the news. We compute these fractions for all the three class variants of the strength of statement (results shown for only the 7-class and the 4-class in Figure~\ref{Figure:Arrival}). An intriguing observation is that, across all the class variants of the strength of statement, the fraction of tweets in the $tweet_{EN}$ category is higher after one year compared to the fraction of tweets in the $tweet_{NEN}$ category. Motivated by this observation, in the following, we analyze the linguistic structure of the \textit{late} tweets (that arrive after one year) and compare them with the \textit{early} tweets (that arrive within one day).

\begin{table}
\centering
\small
\begin{tabular}{|c|c|c|c|c|c|}
\hline
Exagg. class &Tweet type & L & R & H & M \\ \hline
\multirow{2}{*}{7 class } &$tweet_{EN}$ & \cellcolor{green}\textbf{0.06} & \cellcolor{green}\textbf{0.21} & 0.23 & \cellcolor{green}\textbf{0.23} \\
						  &$tweet_{NEN}$ & 0.05 & 0.17 & 0.26 & 0.14 \\ \hline
\multirow{2}{*}{4 class } &$tweet_{EN}$ & \cellcolor{green}\textbf{0.06} & \cellcolor{green}\textbf{0.20} & 0.23 & \cellcolor{green}\textbf{0.23} \\
						  &$tweet_{NEN}$ & 0.05 & 0.18 & 0.26 & 0.15 \\ \hline
\multirow{2}{*}{2 class } &$tweet_{EN}$ & \cellcolor{green}\textbf{0.06} & \cellcolor{green}\textbf{0.20} & 0.23 & \cellcolor{green}\textbf{0.23} \\
						  &$tweet_{NEN}$ & 0.05 & 0.18 & 0.25 & 0.15 \\ \hline

\end{tabular}%
\vspace{0.2cm}
\caption{Fraction of important attributes (per tweet) for the two types of tweets in various classes of exaggeration. L: Likes, R: Retweets, H: Hashtags and M: Mentions.}
\label{Table:Tweet}
\vspace{-0.8cm}
\end{table}

\begin{figure*}[h]
\centering

\subfloat[anger]{\label{Figure:Anger}\includegraphics[width=0.2\textwidth]{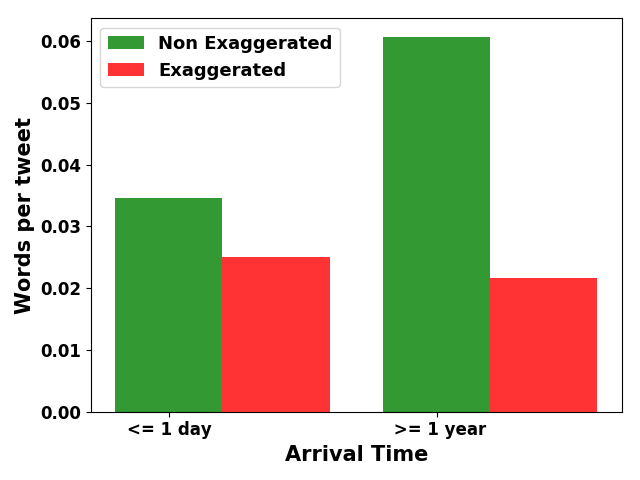}}
\subfloat[negative emotion]{\label{Figure:Negemo}\includegraphics[width=0.2\textwidth]{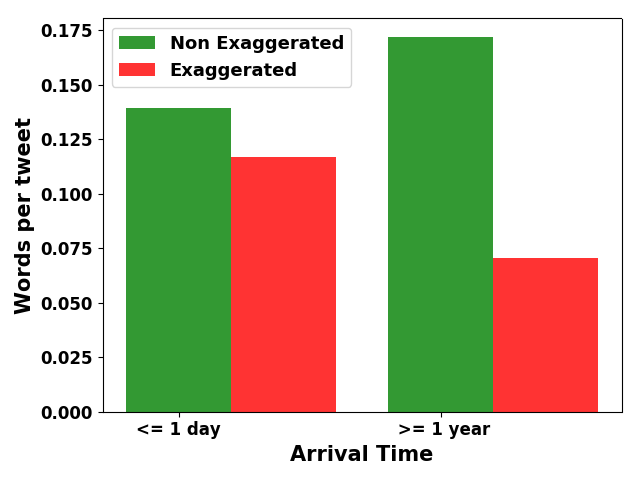}}
\subfloat[ingestion words]{\label{Figure:Ingest}\includegraphics[width=0.2\textwidth]{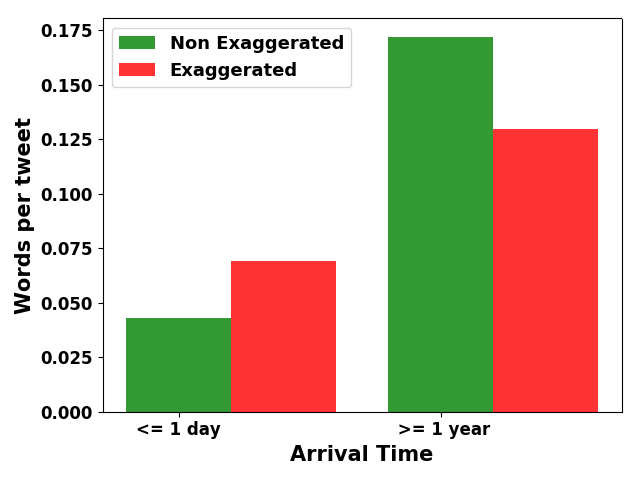}}
\subfloat[leisure]{\label{Figure:Leisure}\includegraphics[width=0.2\textwidth]{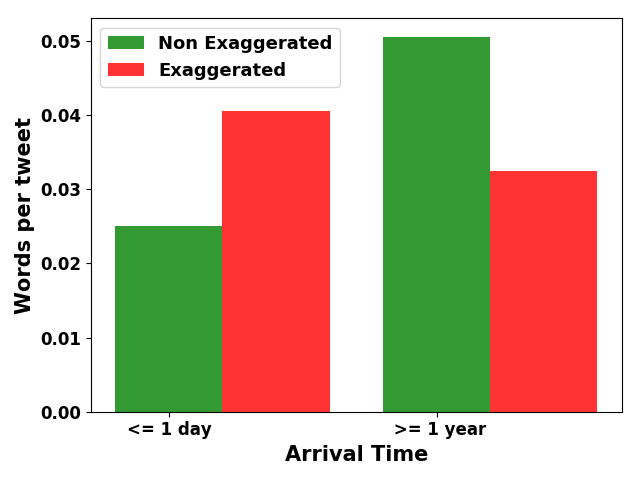}}
\\
\subfloat[sad]{\label{Figure:Sad}\includegraphics[width=0.2\textwidth]{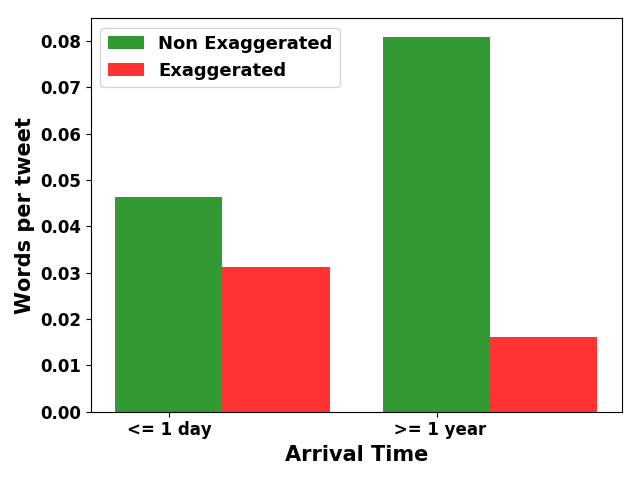}} 
\subfloat[sexual]{\label{Figure:Sexual}\includegraphics[width=0.2\textwidth]{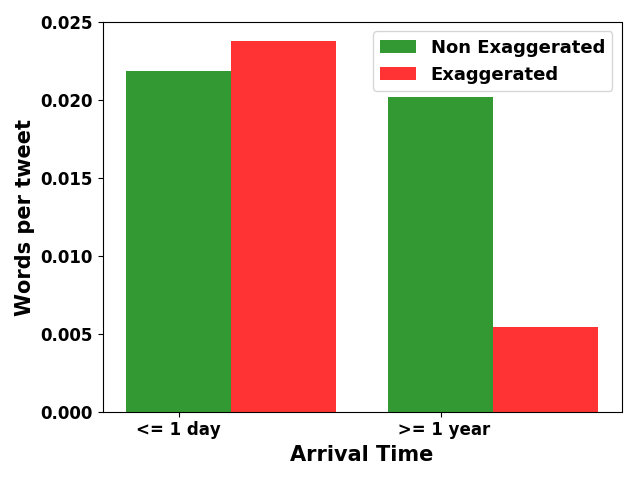}}
\subfloat[death words]{\label{Figure:Death}\includegraphics[width=0.2\textwidth]{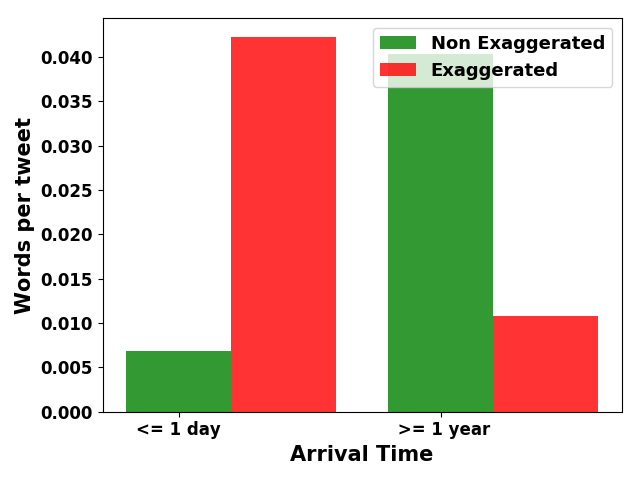}}
\subfloat[swear]{\label{Figure:Swear}\includegraphics[width=0.2\textwidth]{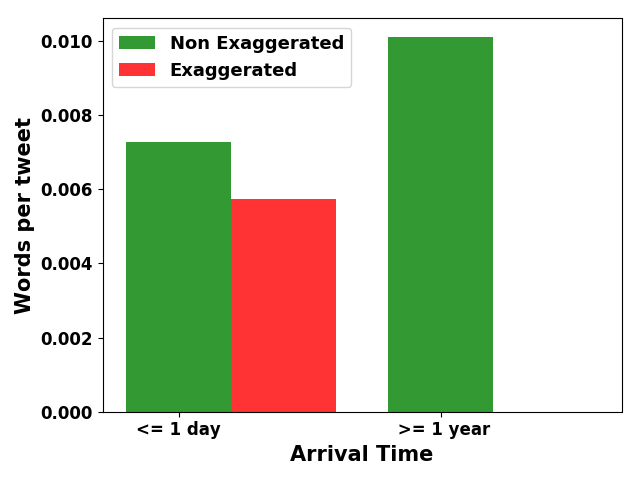}} \\

\caption{LIWC categories having $r^{c}<<1$.}
\label{Figure:LIWCLow}
\end{figure*}

\subsection{Analysis of the tweets}

To obtain a non-trivial set of tweets for the analysis, we filter the tweet text by removing the news headline from the text if it exists, any domain words like \textit{news}, \textit{bbc}, \textit{telegraph} and user mentions (@). Next we perform LIWC (http://www.liwc.net/comparison.php) analysis on the tweet text separately on the early and the late tweets for both the classes. The LIWC dictionary has 64 pre-defined categories (e.g., \textit{pronouns}, \textit{verbs}, \textit{adverbs}, \textit{social}, \textit{family}, \textit{friends} etc.). For each category $c$, we find the number of words per tweet belonging to that category separately from the early and the late tweets of the $tweet_{EN}$ and the $tweet_{NEN}$ groups. Let $eEarly_c$ and $eLate_c$ denote the respective fractions for the early and the late tweets of the $tweet_{EN}$ group in the LIWC category $c$. Similarly, let $neEarly_c$ and $neLate_c$ denote the respective fractions for the early and the late tweets of the $tweet_{NEN}$ group in the LIWC category $c$. We now compute two ratios -- $r^{c}_{EN}=\frac{eLate_c}{eEarly_c}$ and $r^{c}_{NEN}=\frac{neLate_c}{neEarly_c}$

If $r^{c}_{EN} > 1$, it means that the fraction of words of category $c$ has increased in the $tweet_{EN}$ group over time. Similarly, if $r^{c}_{NEN} > 1$ then it means that the fraction of words of category $c$ has increased in the $tweet_{NEN}$ group over time. Finally, we compute the ratio of these two ratios $r^{c}=\frac{r^{c}_{EN}}{r^{c}_{NEN}}$. 

\begin{figure}[!htbp]
\centering

\subfloat[assent]{\label{Figure:Assent}\includegraphics[width=0.16\textwidth]{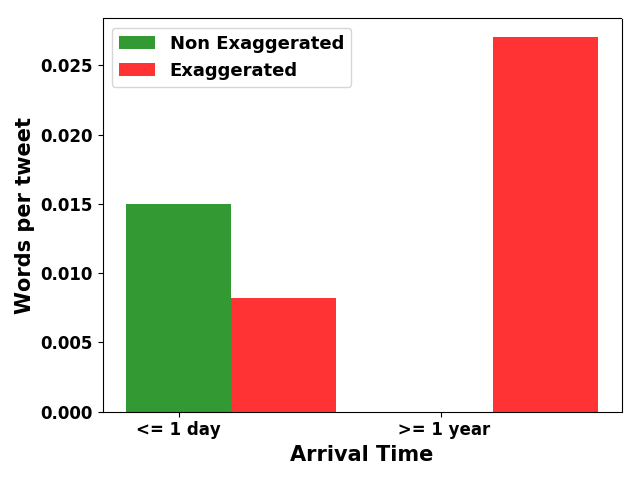}}
\subfloat[future]{\label{Figure:Future}\includegraphics[width=0.16\textwidth]{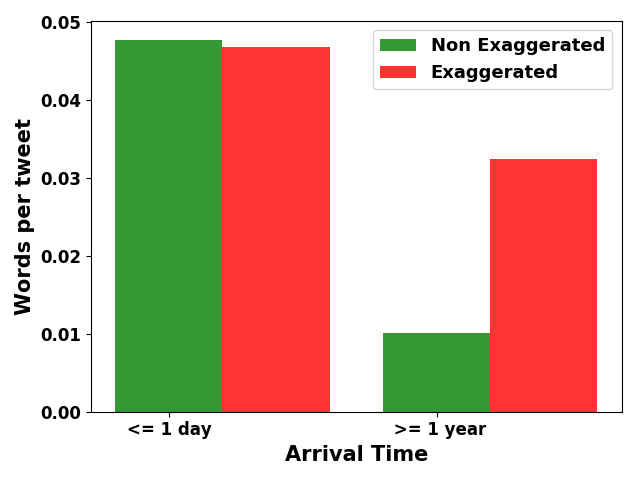}}
\subfloat[feel]{\label{Figure:Feel}\includegraphics[width=0.16\textwidth]{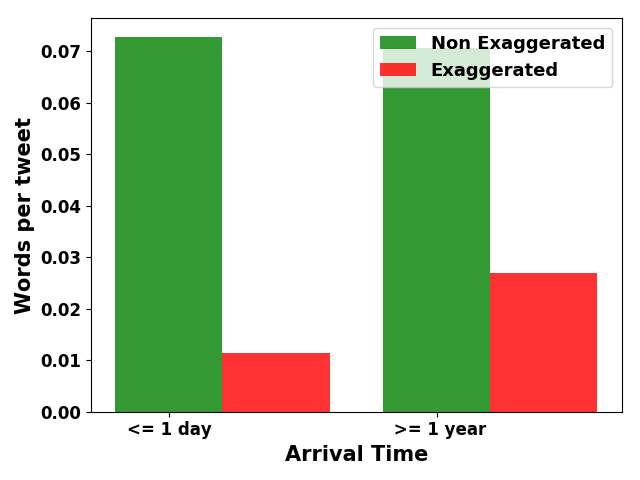}}\\

\caption{LIWC categories having $r^{c}>>1$.}
\label{Figure:LIWCHigh}
\vspace{-0.1cm}
\end{figure}

If $r^{c} > 1$ then the increase, over time, in the fraction of words in category $c$ is more pronounced in the $tweet_{EN}$ group than in the $tweet_{NEN}$ class. We now consider the two extreme ends of this ratio, i.e., cases where either $r^{c} >> 1$ or $r^{c} << 1$. The former signifies stronger presence of the fraction of words in a category $c$ in the late tweets spreading
exaggerated news relative to the late tweets spreading
non-exaggerated news. The latter, on the other hand, indicates a weaker presence of the fraction of words in a category $c$
in the late tweets spreading exaggerated news relative to the late tweets spreading non-exaggerated news. The LIWC categories having extreme low and high values of $r^{c}$ \textcolor{black}{across all the class variants of the strength of statement} are shown in Figures~\ref{Figure:LIWCLow} and~\ref{Figure:LIWCHigh}, respectively. Some immediate observations are as follows. The late tweets spreading exaggerated news (i.e., the $tweet_{EN}$ class) show stark differences in certain LIWC categories -- while the words in categories like \textit{death words}, \textit{sexual}, \textit{sad} and \textit{negative emotion} indicate a weaker presence, a stronger presence is observed in categories like \textit{assent}, \textit{future}, and \textit{feel}. 

\if{0}
Some of the interesting late tweets (representative) from the $tweet_{EN}$ group where $r^{c}>>1$ are as follows. Recall that these are cases where the fraction of words in the category $c$ exhibit a stronger presence in the late tweets that spread exaggerated news ($tweet_{EN}$) relative to the late tweets that spread non-exaggerated news ($tweet_{NEN}$).

\vspace{0.2cm}
\begin{tcolorbox}[boxsep=0pt, top=1pt,left=1pt,right=1pt,bottom=1pt,colframe=gray!50]
\textbf{assent}: \\
{\em $tweet^{late}_{EN}$}:  hmm, i what siri has to say about a diabetes youre sure to smile at this one \textit{URL}
\end{tcolorbox}
\fi
\if{0}
Similarly, some of the interesting late tweets (representative) from the $tweet_{NEN}$ group where $r^{c}<<1$ are as follows. Note that these are cases where the fraction of words in the category $c$ have increased much more in the late tweets that spread non-exaggerated news ($tweet_{NEN}$) compared to the late tweets that spread exaggerated news ($tweet_{EN}$).
\vspace{0.2cm}
\begin{tcolorbox}[boxsep=0pt, top=1pt,left=1pt,right=1pt,bottom=1pt,colframe=gray!50]
\textbf{anger}: \\
{\em $tweet^{late}_{NEN}$}: article from 2011 ecstasy is known to kill some cancer cells, but scientists have increased its effectiveness... \textit{URL}
\end{tcolorbox}

\begin{tcolorbox}[boxsep=0pt, top=1pt,left=1pt,right=1pt,bottom=1pt,colframe=gray!50]
\textbf{negative emotion}: \\
{\em $tweet^{late}_{NEN}$}:  guess thats why you guys are so dumb \textit{URL}
\end{tcolorbox}
\vspace{0.2cm}
\fi
At this point we would like to point out that distinguishing LIWC categories (both $r^{c}<<1$ and $r^{c}>>1$) noted above are consistent across the definitions of exaggeration assuming the 7-class, the 4-class and the 2-class variants of the strength of statement. There are certain categories like \textit{anxiety}, \textit{religion}, \textit{personal pronouns} etc. for which we observed distinctions for the 7-class and the 4-class but not for the 2-class. Deeper investigation revealed that for the 7-class and the 4-class, the number of tweets featuring such words were just one or two thus (falsely) signaling a distinction. This disappears when we consider the 2-class. These words therefore are not actually distinctive and get correctly flagged so because we consider the three class variants instead of a single one. 


\if{0}
\begin{table}[!htbp]
\centering
\begin{tabular}{|c|c|c|c|}
\hline
LIWC CATEGORY & 7-CLASS & 4-CLASS & 2-CLASS \\ \hline
Anxiety Words & 6.067 & 6.067 & 0.627 \\ \hline
See Words & 3.537 & 3.537 & 0.589 \\ \hline
Hear Words & 2.588 & 2.588 & 0.193 \\ \hline
Motion Words & 2.161 & 2.161 & 0.779 \\ \hline
Certain Words & 1.400 & 1.400 & 0.528 \\ \hline
Negations & 1.422 & 1.422 & 0.615 \\ \hline
Personal Pronouns & 1.225 & 1.225 & 0.639 \\ \hline
Prepositions & 0.951 & 0.951 & 1.205 \\ \hline
Biological Words & 0.708 & 0.708 & 1.143 \\ \hline
Tentative Words & 0.912 & 0.912 & 1.341 \\ \hline
Articles & 0.951 & 0.951 & 1.205 \\ \hline
\end{tabular}
\vspace{0.2cm}
\caption{Unstable LIWC categories with $r^c$ values for 7-, 4- and 2-class variants}
\label{Table:Unstable_LIWC}
\vspace{-0.8cm}
\end{table}
\fi
 
\subsection{Important insights from the analysis of tweets} 

Two key insights that we obtain in the course of the LIWC analysis are the following. 

\noindent\textbf{Opinion tweets}: The late tweets that spread exaggerated news show an increase in the proportion of opinionated tweets. Some illustrative examples are 
\vspace{0.2cm}
\begin{tcolorbox}[boxsep=0pt, top=1pt,left=1pt,right=1pt,bottom=1pt,colframe=gray!50]
\noindent something to think about. from news -
\end{tcolorbox}

\begin{tcolorbox}[boxsep=0pt, top=1pt,left=1pt,right=1pt,bottom=1pt,colframe=gray!50]
\noindent - anyone noticed benefits of ?
\end{tcolorbox}

\begin{tcolorbox}[boxsep=0pt, top=1pt,left=1pt,right=1pt,bottom=1pt,colframe=gray!50]
\noindent - should \#midwives undertake this test?
\end{tcolorbox}

\begin{tcolorbox}[boxsep=0pt, top=1pt,left=1pt,right=1pt,bottom=1pt,colframe=gray!50]
\noindent - no genetic testing needed, i know i have this gene!
\end{tcolorbox}

\begin{tcolorbox}[boxsep=0pt, top=1pt,left=1pt,right=1pt,bottom=1pt,colframe=gray!50]
\noindent this is great news if it really works.
\end{tcolorbox}
\vspace{0.2cm}

\begin{figure}[!htbp]
\centering    
\subfloat[opinion words]{\label{Figure:opinion}\includegraphics[width=0.5\linewidth] {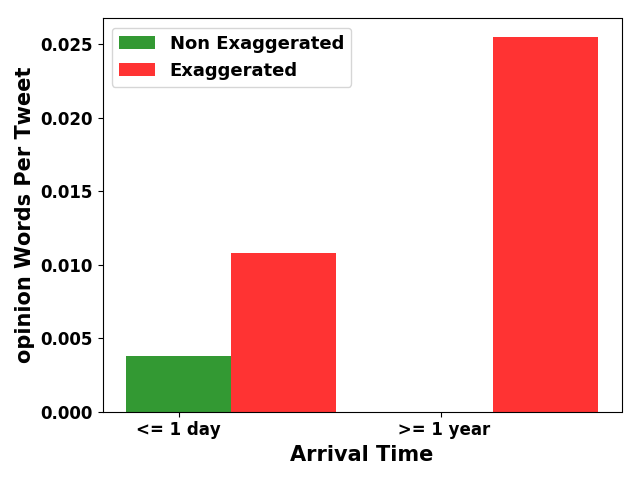}}
\subfloat[realization words]{\label{Figure:realize}\includegraphics[width=0.5\linewidth]{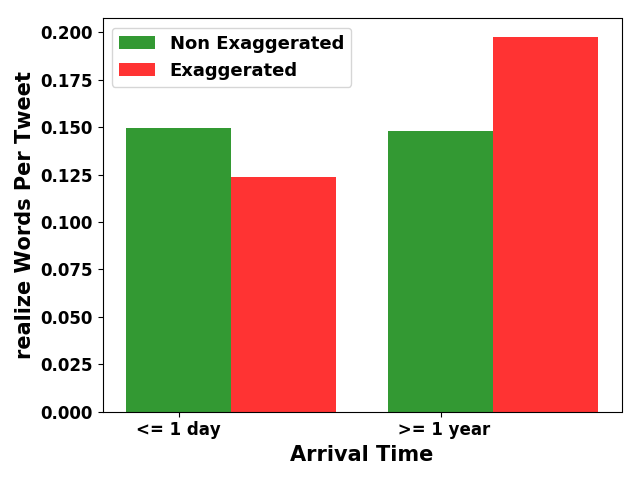}}
\caption{Fraction of opinion and realize words per tweet over time in the $tweet_{EN}$ and $tweet_{NEN}$ classes.}
\end{figure}

This shows that the crowd is not just absorbing and spreading the news blindly but is forming a strong opinion about the same. Tracking these opinions over time therefore can help in developing a completely crowd-sourced mechanism to identify exaggerated news. To validate this further, we create a dictionary of opinion phrases\footnote{Opinion lexicon: \href{https://www.dropbox.com/s/09d21meb3cyb37p/OpinionPhrases?dl=0}{List of Opinion Phrases}} and calculate the number of opinion phrases occurring on average in the early and late tweets of the $tweet_{EN}$ and the $tweet_{NEN}$ classes. As shown in Figure~\ref{Figure:opinion}, the number of opinion phrases increases drastically for the late tweets in the $tweet_{EN}$ class (results are shown when the definition of exaggeration assumes the 7-class variant of the strength of statement; for the 4-class and the 2-class variants the trends are very similar).

\noindent\textbf{Realization tweets}: Another very interesting insight that we gain is that the late tweets that spread exaggerated news also show an increase in the proportion of ``realization'' among the crowd. Some illustrative examples are noted in Table~\ref{Table:Realization} along with the source news headline. This immediately shows that the crowd does not behave like a dumb box; instead the wisdom of the crowd is fully functional and becomes even stronger when exposed for longer time periods. This growing wisdom can be once again leveraged to confirm the authenticity of news articles. To strengthen our point we calculate the fraction of `realize' words per tweet in the early and late tweets of both the $tweet_{EN}$ and $tweet_{NEN}$ classes. We use Pwerthesaurus' realization  lexicon\footnote{Realize lexicon: \url{https://www.powerthesaurus.org/realize}} with a minimum similarity rating of 5 for this purpose. As shown in Figure~\ref{Figure:realize}, the fraction of realization words increases in tweets of the $tweet_{EN}$ class over the time, whereas for the $tweet_{NEN}$ class, this fraction remains the same (results are once again shown when the definition of exaggeration assumes the 7-class variant of the strength of statement; the trends remain very similar for the 4-class and the 2-class variants).

\begin{table*}[!htbp]
\centering
\caption{Tweets demonstrating realization of the crowd.}
\label{Table:Realization}
\footnotesize
\begin{tabular}{|p{8.5cm}|p{8.5cm}|}
\hline
News headline                                                                         & Tweet body                                                                                \\ \hline
Violent homes have the 'same effect on brains of children as combat does on soldiers' & did you know? can same impact that has a soldier?                                       \\
Violent homes have the 'same effect on brains of children as combat does on soldiers' & i might but it depends person who sees it and what happened in your childhood \#appcar          \\
Poor diets may lower children's IQ                                                    & iq. this is child abuse. \#childism                                                       \\
Why the deaf see better than those who can hear                                       & well, for me, i use my glasses just because of my vision doesn't mean i get blind.      \\
Why the deaf see better than those who can hear                                       & do you agreed with scientists discovered that have vision hearing??                     \\
\hline
\end{tabular}
\end{table*}

\section{Users spreading exaggerated news}

In this section we compare the tweeting behavior of the users who post exaggerated news content more frequently with those who never or rarely post exaggerated content.

\noindent\textbf{Dataset}: For this purpose we crawl the timelines of all the users posting the $tweet_{EN}$ and $tweet_{NEN}$ tweets. This results in 8060 unique users and 21,029,228 timeline tweets. Based on the number of exaggerated news in our dataset a user has tweeted about, we divide the users into four categories (i) $users_{NEX}$: set of users who never tweet about any exaggerated news, (ii) $users_{EX1}$: set of users who tweet about exactly one exaggerated news (i.e., rarely), (iii) $users_{EX2}$: set of users who tweet about two exaggerated news, and (iv) $users_{EX3}$: set of users who tweet about three or more exaggerated news. The distribution of the number of users and the volume of timeline tweets across the different user categories are shown in Table~\ref{Table:UserDist}. 

\begin{figure}
\centering    
\subfloat[Average retweet per tweet per user.]{\label{Figure:AvgRt}\includegraphics[width=0.25\textwidth]{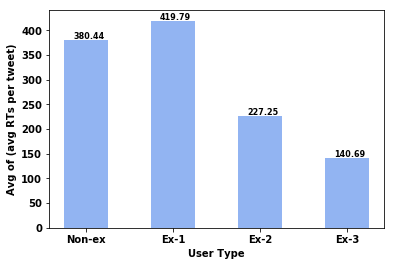}}
\subfloat[Average mention per tweet per user.]{\label{Figure:AvgMen}\includegraphics[width=0.25\textwidth]{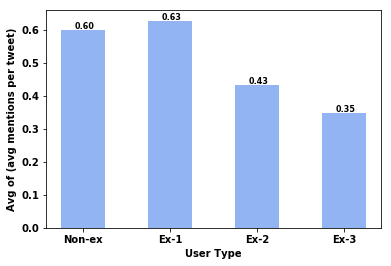}} 
\caption{Average retweet and mention per tweet for each user categories.}
\label{Figure:AvgRtMen}
\end{figure}

\begin{figure}[h]
\centering
\includegraphics[width=0.35\textwidth,height=3.5cm]{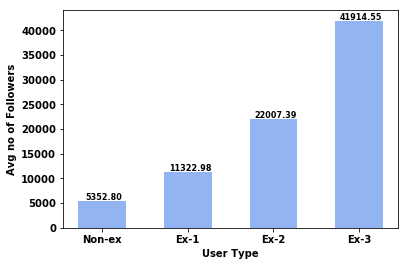}
\caption{Follower count distribution across user categories.}
\label{Figure:AvgFollowercountPerCat}
\end{figure}

\begin{figure*}
\centering    
\subfloat[Slang]{\label{Figure:Slang} \includegraphics[width=0.3\textwidth]{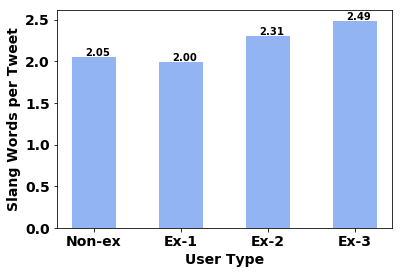}} 
\subfloat[Contraction]{\label{Figure:Contraction} \includegraphics[width=0.3\textwidth]{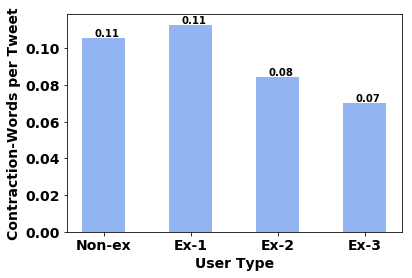}} 
\subfloat[Hyperbolic]{\label{Figure:Hyperbolic} \includegraphics[width=0.3\textwidth]{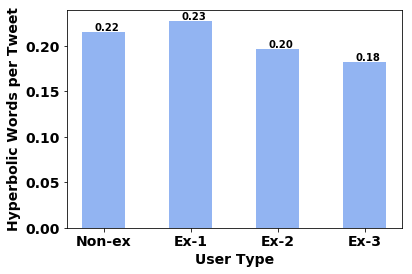}}
\caption{Various linguistic properties of the timeline tweets for various user categories.}
\label{Figure:LangCatrgories}
\end{figure*}

\noindent\textbf{Key results}: Figure~\ref{Figure:AvgRtMen} shows the average number of retweets and mentions that a user gets per tweet in each category. Users who post exaggerated news more frequently tend to get less retweets/mentions. The average number of followers of the users who post more exaggerated content is significantly higher (see Figure~\ref{Figure:AvgFollowercountPerCat}.).
Users who post exaggerated content more frequently use more slang words\footnote{These words include Internet slangs like LOL, LMAO, AMA etc.}, less hyperbolic words\footnote{These are words that have very high positive sentiment e.g., `soul-stirring', `gut-wrenching' etc. See~\cite{Chakraborty:2016} for details.} and less word contractions\footnote{These words typically refer to abbreviations like `they're', `you're', `you'll', `we'd' etc.} in their tweets (see Figure~\ref{Figure:LangCatrgories}). Further LIWC analysis of the timeline tweets shows that the categories like `bio', `health', `body' and `negative emotion' are dominant in case of users who post exaggerated news more frequently.

\noindent\textbf{Prediction model}: We attempt to build a classifier that can automatically distinguish between the two classes of users -- (class I: $users_{NEX}$, $users_{EX1}$), i.e., those who never or rarely post exaggerated content and (class II: $users_{EX2}$, $users_{EX3}$), i.e., those who post exaggerated content more frequently. We use the features described above like the retweet and mention count, count of slang, hyperbolic and contraction words and LIWC features. In addition we also use simple features like the tweet count, the word count, the total word length, the number of stop words and the number of common phrases (commonly used catchphrases adapted from~\cite{Chakraborty:2016}). We do a stratified ten fold cross validation for the evaluation. The precision, recall and F1-scores for different classifiers are presented in Table~\ref{Table:Classification}. Note that we use appropriate class weighting for both classification as well as precision/recall/F1-score calculation since our classes are unbalanced (number of users in class I is much larger than the number of users in class II). 

\begin{table}
\centering
\small
\begin{tabular}{|l|l|l|}
\hline
User type     & $|Users|$ & $|Timeline\_tweets|$ \\ \hline
$users_{NEX}$ &   3,019    &    7,870,784         \\
$users_{EX1}$ &   4,024    &    10,191,495        \\
$users_{EX2}$ &   584      &    1,634,397         \\
$users_{EX3}$ &   433      &    1,332,552         \\
\hline
\end{tabular}
\caption{Distribution of number of users and time-line tweets across user types.}
\label{Table:UserDist}
\end{table}

\if{0}
\begin{figure}
\centering    
\subfloat[Maximum retweet per tweet per user]{\label{Figure:MaxRt}\includegraphics[width=0.25\textwidth]{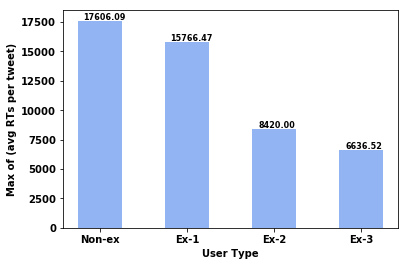}}
\subfloat[Maximum mention per tweet per user]{\label{Figure:MaxMen}\includegraphics[width=0.25\textwidth]{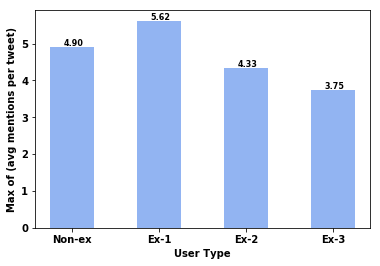}} 
\caption{Maximum of average retweet and mention per tweet for each user categories}
\label{Figure:MaxRtMen}
\end{figure}
\fi

\begin{table}
\centering
\small
\begin{tabular}{|l|l|l|l|}
\hline
Classifier     & Precision & Recall & F1 Score \\ \hline
Naive Bayes    &   0.83    & 0.69  & 0.74     \\
\rowcolor{green}Random Forest  &   0.83    & 0.87  & 0.83     \\
SGD Classifier &   0.72    & 0.43  & 0.64     \\
\rowcolor{green}XGBoost        &   0.83    & 0.87  & 0.83     \\
SVC            &   0.76    & 0.87  & 0.81     \\
\hline
\end{tabular}
\caption{Class weighted precision, recall and F1-score for classification of users into class I and class II.}
\label{Table:Classification}
\end{table}

\section{Conclusion}


In this paper we observe that tweets spreading exaggerated news items show distinctive increase in proportions at later times. While fraction of words in certain LIWC categories like \textit{assent}, \textit{future} etc. show a stronger presence in such tweets, for certain others like \textit{anger} and \textit{negative emotion} there is a weaker presence. A deeper digging into the text further reveals that such tweets are more opinionated and portray larger realization by the crowd. Users posting exaggerated content more frequently are distinct in their tweeting behavior. 

One of the most crucial future steps would be to study the cascading properties of exaggerated content in social media. For instance, it might be interesting to estimate and be able to predict the depth of the spread of exaggerated content over the social network. Another direction would be to develop a framework to automatically detect if a statement is an exaggerated version of the other.  



\bibliographystyle{aaai}
\bibliography{main}
\end{document}